\documentclass{article}

\let\ced\c
\let\lslash\l
\let\Lslash\L
\usepackage{jstyle}
\usepackage{bm}

\newcommand{\arXiv}[1]{\href{http://www.arXiv.org/abs/#1}{arXiv:#1}}
\newcommand{\beq}{\begin{equation}}
\newcommand{\eeq}{\end{equation}}
\newcommand{\del}{\partial}

\makeatletter
\newcommand{\ostar}{\mathbin{\mathpalette\make@circled\star}}
\newcommand{\make@circled}[2]{%
  \ooalign{$\m@th#1\smallbigcirc{#1}$\cr\hidewidth$\m@th#1#2$\hidewidth\cr}%
}
\newcommand{\smallbigcirc}[1]{%
  \vcenter{\hbox{\scalebox{0.77778}{$\m@th#1\bigcirc$}}}%
}
\makeatother

\newtheorem{proposition}{Proposition}
\newtheorem{remark}{Remark}
\newtheorem{theorem}{Theorem}
\newtheorem{lemma}{Lemma}

\def\al{\alpha}
\def\be{\beta}
\def\de{\delta}
\def\eps{\varepsilon}
\def\i{\iota}
\def\s{\star}
\def\w{\wedge}
\def\Tr{\mathrm{Tr}}
\def\sfi{\ostar}
\def\zz{z}

\title{Nonlinear (chiral) $\bm p\hspace{0.5mm}$-form electrodynamics}

\author[a]{Zhirayr Avetisyan,}
\author[b,c]{Oleg Evnin,}
\author[d]{Karapet Mkrtchyan}

\affiliation[a]{Department of Mathematics: Analysis, Logic and Discrete Mathematics,\\ University of Ghent, 9000 Ghent, Belgium}
\affiliation[b]{Department of Physics, Faculty of Science, Chulalongkorn University, Bangkok 10330, Thailand}
\affiliation[c]{Theoretische Natuurkunde, Vrije Universiteit Brussel (VUB) and\\ International Solvay Institutes, Pleinlaan 2, Brussels 1050, Belgium}
\affiliation[d]{Theoretical Physics Group, Blackett Laboratory, Imperial College London SW7 2AZ, U.K.}

\emailAdd{\mbox{jirayrag@gmail.com, oleg.evnin@gmail.com, k.mkrtchyan@imperial.ac.uk}}

\abstract{\vspace{2mm}\\In our previous article Phys.~Rev.~Lett. 127 (2021) 271601, we announced a novel `democratic' Lagran\-gian formulation of general nonlinear electrodynamics in four dimensions that features electric and magnetic potentials on equal footing. 
Here, we give an expanded and more detailed account of this new formalism, 
and then proceed to push it significantly further by building the corresponding Lagrangian theories
of higher form field interactions in arbitrary dimensions.
Special attention is given to interactions of chiral $2k$-forms in $4k+2$ dimensions, with further details for 2-forms in 6 dimensions and 4-forms in 10 dimensions. We comment more broadly on the structure of covariant equations of motion for chiral fields, and on the place of our Lagrangian theories in this context. The Lagrangian theories we develop are simple and explicit, and cover a much broader class of interactions than all past attempts in the literature.}

\begin{document}

{\phantom{.}\vspace{-2.5cm}\\\flushright Imperial-TP-KM-2022-1\\
}
\bigskip

\maketitle

\section{Introduction}

The symmetry between electric and magnetic degrees of freedom received appreciation rather quickly
following the creation of Maxwell's theory of electromagnetism. A comprehensive summary of early developments can be found in \cite{history}. A further boost was given to this topic by Dirac's analysis of magnetic monopoles \cite{dirac}, which spawned the field of monopole phenomenology and the corresponding experimental searches \cite{mnpl}. On the more theoretical front, the ideas of electric-magnetic duality have resurfaced prominently in such developments as the dual monopole condensation picture of confinement \cite{mandelstam,simonov}, Montonen-Olive duality \cite{MO} and the Seiberg-Witten solution of supersymmetric gauge theories \cite{SW}. Related notions have also played a key role in the `second superstring revolution' \cite{2ndstr1,2ndstr2,2ndstr3}.

In the original Maxwell theory, it is impossible to identify the electric and magnetic fields with each other in a Lorentz-covariant way, as that would leave only vanishing field configurations.\footnote{We work in Minkowski spacetimes.} This, however, becomes possible if the number of spacetime dimensions equals 2~mod~4, with Maxwell's fields replaced by the closely analogous Abelian higher form fields. Forcing the electric fields to equal the magnetic ones in those cases halves the dynamical content of the theory and brings in the notion of {\it chiral} forms. Such chiral forms, whose electric and magnetic components are identified, have come to play an important role in higher-dimensional supergravities \cite{iib1,iib2,iib3}, both as fundamental fields and in effective worldvolume theories of extended solitonic solutions (branes) \cite{fivebrane1,fivebrane2}.

While electric-magnetic duality or selfdualify can be easily seen in the equations of motion, at least within the free theory, it is known to be challenging to incorporate them into Lagrangian theories. Thus, writing the simple equation
\beq
dA_1=\pm \s\! dA_2
\label{A1A2}
\eeq
for two vector potentials $A_1$ and $A_2$ (with $\s$ being the Hodge star) automatically implies that 
both potentials satisfy Maxwell's equations $d\s dA_1=d\s dA_2=0$, where $A_1$ can be understood, by (\ref{A1A2}), as the electric potential and $A_2$ as its magnetic dual. Finding a Lagrangian theory that generates (\ref{A1A2})
as its equation of motion is, however, highly nontrivial, even for the elementary example of a free Maxwell field.
Equations of the form (\ref{A1A2}) are often referred to as twisted selfduality relations, since they can be thought of as selfduality under a compound operation that combines the Hodge duality and the interchange $A_1\to A_2$, $A_2\to -A_1$. Formalisms that include explicit electric and magnetic potentials, as the above $A_1$ and $A_2$, are also referred to as `democratic'.

Tension has long been perceived between locality and Lorentz invariance if one tries to construct a Lagrangian description of a (twisted or ordinary) selfdual field \cite{DT,MSch}.
The first well-known attempt in this direction is \cite{Zw}, which provides a local Lagrangian theory
for Maxwell electrodynamics that features both electric and magnetic potentials, but is not manifestly Lorentz-invariant. A considerable volume of work has followed over the subsequent decades generating a large number of approaches to this problem, each with distinctive advantages and disadvantages \cite{Sieg,KMkrtch,FJ,HT,BS,Harada,Tse1,MYW,Wot,Tse2,SchS,KhP,PST1}. (For an incomplete collection of other relevant historical literature see \cite{Gaillard:1981rj,Bialynicki-Birula:1984daz,Gibbons:1995cv,PST2,PSTproc,Tse3,DH,PSch,PST3,Tseytlin:1997csa,S5M,Berman:1997iz,DallAgata:1997gnw,CW,BBS,Nurmagambetov:1998gp,MPS,PSTnotoph,RTse,MMMK,LM,Hatsuda:1999ys,KT,MMMK2,Bekaert:2001wa,Sorokin,Ivanov:2003uj,Bunster:2010kwr,Bunster:2011qp,non5brane,PST4,BH,KSV,Ivanov:2014nya,Hiroshi,Sen1,Samtleben:2011eb,Bandos:2013jva,Ferrara:1997gh}, for a sampler of more recent related works see \cite{AEShJ,Sen2,BLM1,Buratti:2019guq,Lambert:2019diy,Townsend1,Townsend2,Andriolo:2020ykk,Henneaux:2020nxi,BLST1,Bertrand:2020nob,Kosyakov,Vanichchapongjaroen,Bandos:2020hgy,Cremonini:2020skt,Svazas:2021ltf}.)

Our purpose in this article is to present a general formalism that is simple, economical and accommodates both democratic description of interacting $p$-forms in any number of dimensions, and chiral forms in those dimensions where they exist. This formalism builds upon the previous considerations reported in a series of recent works involving the present authors \cite{Mkrtchyan:2019opf,polynom,democ}. Of the broad variety of approaches pursued in the past literature, this formalism shows closest affinity to the Pasti-Sorokin-Tonin (PST) formulations \cite{PST1,PST2,PST3}. Indeed, for the free theory \cite{Mkrtchyan:2019opf,polynom}, the corresponding PST representation can be recovered from the Lagrangian theories of \cite{Mkrtchyan:2019opf,polynom} by integrating out
a subset of auxiliary fields. An advantage of the free theories of \cite{Mkrtchyan:2019opf, polynom} is, however, that their
Lagrangians are polynomial (in addition to being local and manifestly Lorentz invariant) and suggest a natural generalization to interacting cases, which is our main target in this work. Once interactions have been introduced, the relation between our theories and PST-type formulations becomes less obvious, though the general structure of the formalism (with a large set of gauge symmetries completely eliminating the auxiliary fields on-shell) remains similar. The key achievement of this paper is that we report a much bigger class of Lagrangian interacting chiral form theories than what has been attained in the past using all other approaches. 

We briefly summarize here one of the main results of this work. A large class of very general equations of motion for a self-interacting Abelian selfdual $p=2k$-form in $d=4k+2$ dimensional Minkowski spacetime is given by the following covariant equations:
\begin{align}\label{FFf}
    F-\star F=f(F+\star F)\,,
\end{align}
where $f:\Lambda^+\to \Lambda^-$ is a function of selfdual $(p+1)$-form variable taking values in anti-selfdual $(p+1)$-forms.\footnote{The same statements are (obviously) true for anti-selfdual fields, with plus and minus signs interchanged.} Our Lagrangian formulation allows for a description of any such theory, 
provided that
\begin{align}\label{fHg}
    f(\mathcal{H})=\frac{\partial \mathcal{F}(\mathcal{H})}{\partial \mathcal{H}}\,,
\end{align}
where $\mathcal{F}:\Lambda^+\to \mathbb R$ is an arbitrary scalar function of a selfdual variable that enters the Lagrangian given by
\begin{align}\label{LFQ}
    \mathcal{L}=(F+aQ)^2+2\,a\,F\wedge Q+\mathcal{F}(\mathcal{H})\,,\qquad \mathcal{H}\equiv (F+aQ)+\star (F+aQ)\,.
\end{align}
Here, $F=dA$ is the field strength of a dynamical $p$-form field $A$, while $Q=dR$ is the field strength of an auxiliary $p$-form field $R$, and $a$ is an auxiliary scalar. The latter two fields are completely eliminated on-shell due to a large set of gauge symmetries of (\ref{LFQ}). In the particular case of six spacetime dimensions, the self-interacting theory of a single chiral two-form is defined by a function of one variable (there is a single functionally independent invariant constructed from a selfdual three-form), and any theory of the very general form (\ref{FFf}) can in fact be derived from a Lagrangian of the form (\ref{LFQ}) since (\ref{fHg}) is automatically satisfied in this case due to the very restricted structure of Lorentz-covariant chiral form functions.

Our exposition is organized as follows: We shall start by reviewing, in section~\ref{nled}, the considerations of \cite{democ}, providing some pedagogical and technical details that had been omitted from the letter-format paper. We then explain, in section~\ref{democp}, how to generalize the interacting theories of \cite{democ} to a democratic formulation of interacting form fields of arbitrary rank in an arbitrary number of dimensions. (This generalization has been briefly alluded to already in the conclusions of \cite{democ}.) In section~\ref{chiralf}, we specialize to the case of chiral $2k$-forms in $4k+2$ dimensions, where we provide explicit details for 2-forms in 6 dimensions and 4-forms in 10 dimensions, the latter case known to be essentially inaccessible to the previously established approaches.\footnote{{We thank Dmitri Sorokin for correspondence on this matter.}}
Given that our consideration produces very simple and explicit equations of motion for the propagating degrees of freedom, we dedicate section~\ref{eoms} to discussing the place these equations occupy relatively to the most general covariant selfduality relation one can write down, as well as to the past approaches, such as \cite{PSch}, where manifest Lorentz covariance is sacrificed in the treatment of these equations. Sections~\ref{chiralf} and \ref{eoms} can be read largely independently from sections~\ref{nled} and \ref{democp}, and this may be a valid strategy for readers specifically interested in chiral form fields. We review the implications of our results in section~\ref{concl}.

\section{Nonlinear electrodynamics in 4 dimensions}\label{nled}

We start with reviewing the considerations of \cite{democ} and the democratic Lagrangian representation of general nonlinear electrodynamics developed there. In this review, we will follow the original logic of \cite{democ} so as to make the underlying heuristics clearly visible to the reader. As already remarked in the conclusions of \cite{democ}, the final structure that emerges from this analysis can be understood in simpler terms, and it is this angle that allows for easy generalization to other field ranks and numbers of dimensions. These generalizations will be the main subject of the rest of the paper. (For readers specifically interested in chiral forms, it should be largely possible to proceed directly to section~\ref{chiralf}.)

A comprehensive contemporary review of nonlinear electrodynamics from a perspective close to our considerations can be found in \cite{intro}.

\subsection{Free theory}

Before proceeding to include interactions, it is wise to recall in some detail the corresponding democratic construction for a free Maxwell field \cite{Mkrtchyan:2019opf,polynom}.

It is a common feature of all approaches to democratic Lagrangians that a set of auxiliary fields is present, and they are eliminated on-shell due to a large set of gauge symmetries, leaving the desired physical degrees of freedom (see, e.g., \cite{Sieg,MYW,PST1,Samtleben:2011eb,Bandos:2013jva}). It is commonly known to be impossible to formulate in a Lagrangian language such democratic theories with explicit electric and magnetic degrees of freedom, or chiral form theories, without including auxiliary fields (see, e.g., \cite{Sieg,MYW,PST1,PST2,PST3}) or sacrificing manifest Lorentz symmetry (see, e.g., \cite{MSch,HT,SchS}).

The set of fields used in all of the constructions in this paper is of the same type: it includes the original `electric' $p$-form field in $d$ spacetime dimensions (Lorentzian signature), its `magnetic' dual $(d-p-2)$-form field (the latter identified with the `electric' field for the chiral cases), an auxiliary form field sector that copies the electric-magnetic sector we have just described, and an extra scalar (whose role is very similar to the auxiliary scalar in PST theories \cite{PST1,PST2,PST3}). 

For the case of 4-dimensional electrodynamics considered in this section, the above specifications translate into an electric-magnetic doublet of vector potentials $A^b_\mu$ with $b=1,2$ being the duality space index (``1'' can be understood as `electric,' and ``2'' as `magnetic') and $\mu$ the ordinary spacetime index, and additionally an auxiliary doublet of vector potentials $R^b_\mu$ and an auxiliary scalar $a$.
The Lagrangian is\footnote{The difference in the sign of the second term compared to \cite{democ} fixes a minor inconsistency.}
\begin{equation}
{\cal L}_{Maxwell}=-\frac14 H^b_{\mu\nu}H^b{}^{\mu\nu}{-}\frac{a}4\,\,\epsilon^{bc}\,\varepsilon^{\mu\nu\lambda\rho}\,F^{b}_{\mu\nu}\,Q^{c}_{\lambda\rho}\,,  \label{Maxwell}
\end{equation}
where $\epsilon$ and $\varepsilon$ are correspondingly two- and four-dimensional Levi-Civita symbols, and $H^b_{\mu\nu}\equiv F^b_{\mu\nu}+a\,Q^b_{\mu\nu}$ with
\begin{equation}
F^b_{\mu\nu}=\partial_{\mu}\,A^b_{\nu}-\partial_{\nu}\,A^b_{\mu}\,,\quad Q^b_{\mu\nu}=\partial_{\mu}\,R^b_{\nu}-\partial_{\nu}\,R^b_{\mu}\,.
\end{equation}
It may seem surprising that this theory is equivalent to a single propagating Maxwell field, 
and the underlying reason is in the large set of gauge symmetries of (\ref{Maxwell}). Besides the usual gradient shifts of the vector potentials (the Lagrangian only depends on the corresponding field strengths $F$ and $Q$) with parameters in the form of arbitrary scalar doublets $u^b(x)$ and $v^b(x)$, 
\begin{align}
&\de a=0,\quad \de A^b_\mu=\del_\mu u^b,\quad \de R^b_\mu=0,\label{symmAgrad}\\
&\de a=0,\quad \de A^b_\mu=0,\quad \de R^b_\mu=\del_\mu v^b,\label{symmRgrad}
\end{align}
the Lagrangian is invariant under two extra symmetries:
\beq
\delta a=0,\quad \delta A^b_\mu= -a u^b \,\del_\mu a,\quad \delta R^b_\mu=u^b\, \del_\mu a, 
\label{dashift}
\eeq
and
\beq
    \delta a=\varphi\,,\quad \delta A^b_{\mu}=-a\,\delta R^b_{\mu},\quad \delta R^b_{\mu}=-\frac{\varphi}{(\partial a)^2}\,\partial^\nu a\,(Q^b_{\nu\mu}-\epsilon^{bc}\s Q^c_{\nu\mu})\,,
\label{symashift}
\eeq
where $\varphi(x)$ in the last line is an arbitrary scalar field parameter.
The Hodge star operation $\s$ that is, of course, central to the topic of electric-magnetic dualities is defined for 2-forms in 4 spacetime dimensions as
\beq
\s Q_{\mu\nu}=\frac12 \varepsilon_{\mu\nu\sigma\rho} Q^{\sigma\rho}.
\eeq

The equations of motion corresponding to (\ref{Maxwell}) are
\begin{align}
&d[\s H^b{-}a\,\epsilon^{bc}\,Q^c]=0\,,\hspace{3mm}\label{Eq:Afree}\\
&d[a(\,\s H^b{+}\epsilon^{bc}\,F^c)]=0\,,\hspace{3mm}\label{Eq:Rfree}\\
&Q^b\w(\s H^b{+}\epsilon^{bc}\,H^c)=0\label{Eq:auxfree},
\end{align}
where we have switched to the differential form notation. (A summary of conventions, identities and conversion formulas related to the tensor and differential form notation can be found in the appendices of \cite{polynom}.) Multiplying (\ref{Eq:Afree}) with $a$ and subtracting it from (\ref{Eq:Rfree}), we obtain
\beq
da\w (\s H^b{+}\epsilon^{bc}\,H^c)=0.
\label{dafree}
\eeq
Taking the Hodge dual of this formula, using the identities $\s v\w=\i_v\s$ (valid for any one-form $v$ with $\i_v$ being the inner product) and $\s\s=-1$, and contracting with $\epsilon^{bd}$ (followed by renaming $d$ to $b$), we obtain
\beq
i_{da} (\s H^b{+}\epsilon^{bc}\,H^c)=0.
\label{idafree}
\eeq
Then, acting with $\i_{da}$ on (\ref{dafree}) and with $da\w$ on (\ref{idafree}), adding up the result, and using the projection-rejection identity valid for any (non-null) 1-form $v$,
\beq
\frac{1}{v^2}\left(v\w\i_v+\i_v \;v\w\right)=1,
\eeq
we deduce that
\beq
\s\! H^b{+}\epsilon^{bc}\,H^c=0
\label{Hself}
\eeq
whenever (\ref{Eq:Afree}) and (\ref{Eq:Rfree}) are satisfied. In particular, it means that the equation of motion for $a$ given by (\ref{Eq:auxfree}) is implied by the remaining equations of motion. This structure is closely related to $a$ being a pure gauge degree of freedom, as the symmetry transformations (\ref{symashift}) imply that it can be changed arbitrarily. Note that $a$ cannot be gauged away completely, and permissible gauges respect the condition $(\del a)^2\ne 0$ at all spacetime points. One indication of this is the presence of $(\del a)^2$ in the denominator in the gauge transformation (\ref{symashift}). 

Finally, substituting (\ref{Hself}) into (\ref{Eq:Afree}), we deduce that
\beq
da\w dR^b=0.
\eeq
Such equations for differential forms are ubiquitous in the approach to (twisted) selfduality we describe here, and they are also very common in the PST theory. A detailed integration procedure can be found in Appendix C of \cite{polynom}. The most general solution is
\beq
R^b=dC^b+da\w E^b,\quad\mbox{or in components,}\quad R^b_\mu=\del_\mu C^b +E^b\del_\mu a \,.
\label{Rraw}
\eeq
However, any such solution can be gauge-transformed to $R^b=0$. Namely, the first term can always be removed by (\ref{symmRgrad}), while the second term can be removed by (\ref{dashift}). Then, with $R^b=0$, (\ref{Hself}) turns into
\beq
\s\! F^b{+}\epsilon^{bc}\,F^c=0,
\label{Fself}
\eeq
which is the same as $F^1= \s F^2$. Applying an exterior derivative to this relation, we conclude that $F^2$ satisfies Maxwell's equation $d\s F^2=0$, while $F^1$ is its magnetic dual. Thus, the only propagating degrees of freedom of (\ref{Maxwell}) are those of an ordinary Maxwell field, while $A^1$ and $A^2$ explicitly present in the formulation of the theory are its electric and magnetic potentials satisfying (\ref{A1A2}).

\subsection{Interactions}

We wish to construct a nonlinear generalization of (\ref{Maxwell}), and a natural strategy is to look for deformations of that theory that preserve its rich gauge symmetry structure. Indeed, it is precisely the gauge symmetries (\ref{symmAgrad}-\ref{symashift}) that ensured that the equation of motion for $a$ is automatically satisfied and that $a$ is a pure gauge degree of freedom, and furthermore allowed us to nullify the $R$-fields in any solution of the equations of motion under (\ref{Rraw}). If similar symmetries exist in a nonlinear deformation of (\ref{Maxwell}), it is natural to expect (and this expectation will hold true, as we are about to show) that they will similarly eliminate the auxiliary fields and leave behind the nonlinear electrodynamics of a single vector field (and its magnetic dual).

A key observation in relation to (\ref{symmRgrad}-\ref{symashift}) that has served as a basis for the derivations in \cite{democ} is that $H^b_{\mu\nu}$ is by itself invariant under (\ref{symmAgrad}-\ref{dashift}), while the last term in (\ref{Maxwell}) changes under the action of (\ref{dashift}) by a total derivative. As a result, the symmetry of the Lagrangian with respect to (\ref{symmAgrad}-\ref{dashift}), as well as the Lorentz transformations, will remain intact if one replaces the first term in (\ref{Maxwell}), namely the term proportional to $H^b_{\mu\nu} H^{b\,\mu\nu}$, by {\it any} scalar made of $H^b_{\mu\nu}$. For a Lagrangian constructed in this way, one must only worry about enforcing the last remaining symmetry (\ref{symashift}) that shifts $a$, and we shall see below that there is a simple way to enforce this symmetry. (It will also become apparent later that one could devise a natural prescription
to implement the last remaining symmetry (\ref{symashift}) automatically from the onset, and this prescription also generalizes naturally to all spacetime dimensions; these considerations will form the core of the subsequent sections.)

Starting with $H^b_{\mu\nu}$, one can build six functionally independent scalars: 
\begin{equation}
    U^{ab}\equiv\frac12\,H_{\mu\nu}^a\,H^b{}^{\mu\nu}\,,\quad V^{ab}\equiv\frac12\,H_{\mu\nu}^a\,\s{H}^{b}{}^{\mu\nu}\,,\label{UV}
\end{equation}
where $U^{ab}=U^{ba}$, $V^{ab}=V^{ba}$. As indicated in the previous passage, we start with the following nonlinear generalization of (\ref{Maxwell}):
\begin{equation}
    {\cal L}= {-} a\,\epsilon_{bc}F^b\wedge Q^c+f(U,^{\hspace{-0.7mm}11}U,^{\hspace{-0.7mm}12}U,^{\hspace{-0.7mm}22}V,^{\hspace{-0.5mm}11}V,^{\hspace{-0.5mm}12}V^{22})\,,\label{NLE}
\end{equation}
which automatically respects the symmetries (\ref{symmAgrad}-\ref{dashift}) of the free theory, but does not automatically respect (\ref{symashift}) until we introduce further constraints on $f$ below.

The equations of motion for $A^b$ and $R^b$ are
\begin{eqnarray}
&d[(f^U_{bc}+f^U_{cb})\,\s H^c-(f^V_{bc}+f^V_{cb})\,H^c+a\,\epsilon_{bc}\,Q^c]=0\,,\hspace{3mm}\label{Eq:A}\\
&d[a\{(f^U_{bc}+f^U_{cb})\,\s H^c-(f^V_{bc}+f^V_{cb})\,H^c-\epsilon_{bc}\,F^c\}]=0\,.\label{Eq:R}
\end{eqnarray}
where $f^U_{ab}\equiv\del f/\del U_{ab}$, $f^V_{ab}\equiv\del f/\del V_{ab}$ ($f^U_{21}\equiv 0\equiv f^V_{21}$).
The equation of motion for $a$ is
\beq
Q^b\wedge\,K_b=0\,,\label{Eq:a}
\eeq
where
\beq
    K_a\equiv(f^U_{ab}+f^U_{ba})\,\s H^b-(f^V_{ab}+f^V_{ba})\,H^b-\epsilon_{ab}\,H^b\,.\label{conjecture4d}
\eeq
In direct analogy to the free theory derivation above (\ref{dafree}), we can multiply
(\ref{Eq:A}) with $a$ and subtract it from (\ref{Eq:R}) to obtain
\beq
    da\wedge K_b=0\,.\label{Kda}
\eeq
We would like to ensure that this equation implies the equation of motion for $a$ (\ref{Eq:a}) in a manner similar to the free theory. This would ascertain that $a$ is a pure gauge degree of freedom. More importantly, it will allow us to derive first-order equations from a second order Lagrangian we started with.
This would always happen, in particular, if
\beq
    K_a+ \epsilon_{ab}\,\s K_b\equiv 0\,\label{Ksd}
\eeq
is satisfied identically. If it is, then (\ref{Kda}) will imply $K_b=0$ in direct parallel to the free theory derivation under (\ref{dafree}). Then, (\ref{Eq:a}) is a consequence of the equations of motion for the form fields (\ref{Eq:A}-\ref{Eq:R}). Admittedly, the described approach is a particular way one could technically enforce the desired structure of the equations of motion, but as we shall see below, it is sufficient to capture generic nonlinear electrodynamics. More broadly, the theories we shall build according to the same principles in the next sections capture a huge class of form field interactions (quite possibly, all that exist).

One can equivalently recast (\ref{Ksd}) as the following condition in terms of $f$:
\beq
    \,\delta^{ac}\,(f^U_{cb}+f^U_{bc})-\epsilon^{ac}\,(f^V_{cb}+f^V_{bc})+\delta^a_b=0\,.\label{MatrixEq}
\eeq
The general solution of these linear PDEs is
\begin{eqnarray}
    &f(U,V)=-\frac12 U_{aa} + g(\lambda_1,\lambda_2)\,,\label{solutionNLE}\\
    &\lambda_1= U_{12}-\frac12\,(V_{11}-V_{22})\,,\label{lambda1}\\
    &\lambda_2=V_{12}+ \frac12\,(U_{11}-U_{22})\,.\label{lambda2}
\end{eqnarray}
where $g(\lambda_1,\lambda_2)$ is an arbitrary function of two variables. The first term in $f$ simply corresponds to the free theory (\ref{Maxwell}) and the resulting full Lagrangian is 
\beq
    {\cal L}={\cal L}_{Maxwell}+g(\lambda_1,\lambda_2)\,.\label{NLED}
\eeq
We note that, using (\ref{lambda1}-\ref{lambda2}) and (\ref{UV}), $\lambda_{1,2}$ can be expressed as
\beq
\lambda_1=\tfrac12\,\mathcal{H}_{\mu\nu}\,\s \!\mathcal{H}^{\mu\nu}\,,\qquad
\lambda_2=-\tfrac12\,\mathcal{H}_{\mu\nu}\,\mathcal{H}^{\mu\nu}\,,
\label{lamH}
\eeq
with $\mathcal{H}_{\mu\nu}\equiv \s H^1_{\mu\nu}-H^2_{\mu\nu}$. It is an important property that, while $H^b$ is invariant under (\ref{symmAgrad}-\ref{dashift}), but not under (\ref{symashift}),
$\mathcal{H}$ is invariant under all the gauge symmetries (\ref{symmAgrad}-\ref{symashift}) of the free theory. As a result, $\lambda_{1,2}$ as well as the full Lagrangian (\ref{NLED}) are invariant under exactly the same set of gauge transformations as the free theory. We shall return to this property (and also demonstrate it explicitly) in the next sections, where it will form the basis for our generalization of the present formalism to higher dimensions. It is a rather remarkable, and very convenient, feature of the present formulation that the gauge symmetries (\ref{symmAgrad}-\ref{symashift}) are universal and do not need to be adjusted depending on the specific interactions one chooses to consider. (Even though the type of interactions we consider here are those that do not deform the free gauge symmetry also in the single-potential formulation, it does not straightforwardly follow that there should be no deformation in the current formulation with its much larger set of gauge symmetries.)

\subsection{Equations of motion and relation to the single-field  formulation}\label{singlef}

For any theory of the form (\ref{NLE}), the equations of motion (\ref{Eq:A}-\ref{Eq:R}) imply (\ref{Kda}).
Then, our class of theories (\ref{NLED}) has been constructed precisely in such a way as to make (\ref{Ksd}) identically satisfied, and then (\ref{Kda}) implies
\beq
    K_b=0\label{K}
\eeq
by an argument identical to the one given under (\ref{dafree}).
Substituting \eqref{K} into \eqref{Eq:A}, one gets
\beq
    da\wedge dR^b=0\,.
\eeq
Thus, the auxiliary forms satisfy exactly the same equation as in the free theory, solved in full generality by (\ref{Rraw}), so that the result can always be gauge-transformed to
\beq
    R^b=0\,,\qquad H^b=F^b\,.\label{R0}
\eeq
Note that, since $K^b$ satisfies \eqref{Ksd}, (\ref{K}) encodes a single independent equation that can be written as
\beq
\s\! F^1+F^2=\,g_2\,(\s F^1-F^2)-\,g_1\,\s(\s F^1-F^2)\,,\label{FF}
\eeq
with 
\beq
g_1\equiv\frac{\del g}{\del\lambda_1},\qquad g_2\equiv\frac{\del g}{\del\lambda_2}.
\label{g1g2def}
\eeq
One can resolve (\ref{FF}) with respect to $F_1$, and by Lorentz invariance (see, e.g., \cite{BLM1}), one must obtain a relation of the form 
\beq
    F^1=\alpha(s,p) F^2 + \beta(s,p) \s F^2\,,\label{F1F2}
\eeq
where $s$ and $p$ are the two independent Lorentz scalars made of $F^2$,
\beq
   s=\frac12 F^2_{\mu\nu}F^2{}^{\mu\nu}\,,\quad p=\frac12 F^2_{\mu\nu}\s F^2{}^{\mu\nu}.
\label{spdef}
\eeq
The specific form of the functions $\alpha$ and $\beta$ would have to be determined on a case-by-case basis by solving the system of algebraic equations given by (\ref{FF}). What is important for us is that only one of the two field strength $F^1$ and $F^2$ is independent due to the nonlinear selfduality relation (\ref{F1F2}). (The linear version of this relation corresponding to the free theory arises from choosing $g=0$, $\al=0$, $\be=1$.)
Thus, all the auxiliary fields have been eliminated by gauge transformation, and there is only one independent propagating gauge field, satisfying a nonlinear twisted selfduality relation.

To complete the picture, it is useful to clarify how the dynamics of the theory (\ref{NLED}) expressed by the twisted selfduality relation (\ref{F1F2})
corresponds to the usual single-field formulation of nonlinear electrodynamics defined by the Lagrangian
\beq
{\cal L}={\cal L}(s,p),
\label{Lsp}
\eeq
which is an arbitrary function of the two invariants (\ref{spdef}), and we have retained the name $A^2$ for the gauge field. (Note that we do not assume any duality invariance at this point, and we will return to theories invariant under duality rotations later.)
The equation of motion of this theory is
\beq
d\left[-\frac{\del{\cal L}}{\del p} F^2+\frac{\del{\cal L}}{\del s} \s F^2\right]=0,
\eeq
where once again, $F^2$ is simply the field strength 2-form corresponding to the gauge potential $A^2$.
Exactly the same equation arises from applying the exterior derivative operator to (\ref{F1F2}), provided that one identifies
\beq
    \alpha(s,p)=-\frac{\partial{\cal L}}{\partial p}\,,\qquad \beta(s,p)=\frac{\partial{\cal L}}{\partial s}\,.
    \label{lagdev}
\eeq
Thus, the theories described by (\ref{Lsp}) and (\ref{NLED}) are dynamically equivalent, given an appropriate identification of the functions appearing in the two Lagrangians.

We can work out the relation between the theories (\ref{Lsp}) and (\ref{NLED}) more explicitly by rewriting (\ref{F1F2}) in term of the invariants as
\beq
    g_1=\frac{2\,\alpha}{\alpha^2+(\beta+1)^2}\,,\qquad g_2=\frac{\alpha^2+\beta^2-1}{\alpha^2+(\beta+1)^2}\,,\label{galphabeta}
\eeq
where $g_1$ and $g_2$ are the derivatives (\ref{g1g2def}).
While $g_{1,2}$ are defined as functions of $\lambda_1$ and $\lambda_2$, they can be expressed in terms of $\al$, $\be$, $s$ and $p$ using the following relations 
\beq
\begin{split}
    \lambda_1=2\,\alpha\,(1+\beta)\,s-[\alpha^2-(1+\beta)^2]\,p\,,\\
\lambda_2=[\alpha^2-(1+\beta)^2]\,s+2\,\alpha\,(1+\beta)\,p\,\label{Lambdas}
\end{split}
\eeq
that can be obtained from (\ref{F1F2}) and (\ref{lambda1}-\ref{lambda2}), while
\beq
    w\equiv\sqrt{\lambda_1^2+\lambda_2^2}=(\alpha^2+(\beta+1)^2)\,\sqrt{s^2+p^2}\,.\label{walphabetasp}
\eeq

If one starts with a democratic theory of the form (\ref{NLED}), $g(\lambda_1,\lambda_2)$ is given, and (\ref{galphabeta}) should be understood as a $2\times 2$ system of nonlinear algebraic equations for $\alpha$ and $\beta$ as functions of $s$ and $p$. Then, from $\al$ and $\be$, one can recover the single-field Lagrangian $\cal L$ by integrating (\ref{lagdev}). It is not obvious at first sight why $\al$ and $\be$ derived in this manner will satisfy the compatibility condition $\del\al/\del s+\del \be/\del p=0$, but that is indeed the case, as we prove in Appendix~\ref{app:conv}.

Similarly, if one starts with a single-field Lagrangian of the form (\ref{Lsp}), one should first obtain $\alpha$ and $\beta$ from (\ref{lagdev}).
Then, (\ref{Lambdas}) provide a $2\times 2$ system of nonlinear algebraic equations to express $s$ and $p$ through $\lambda_1$ and $\lambda_2$. Finally, (\ref{galphabeta}) gives $g_1$ and $g_2$ as functions of $\lambda_1$ and $\lambda_2$. These derivatives should then be integrated to obtain $g$, and again, we provide existence results for this integration procedure in Appendix~\ref{app:conv}.

\subsection{Duality symmetry and conformal invariance} 

The most basic duality symmetry present in the free Maxwell symmetry is the
$Z_4$ interchange of the electric and magnetic potentials $A^1\to A^2$, $A^2\to-A^1$.
The corresponding democratic formulation (\ref{Maxwell}) similarly respects the transformations $H^1\to H^2\,, H^2\to -H^1$, 
which induce $\lambda_1\to -\lambda_1, \lambda_2\to -\lambda_2$ by (\ref{lamH}).
Thus, to maintain this duality symmetry in the interacting theory (\ref{NLED}) one must simply restrict $g(\lambda_1,\lambda_2)$ to satisfy the condition
\beq
    g(-\lambda_1,-\lambda_2)=g(\lambda_1,\lambda_2)\,.\label{DiscreteDuality}
\eeq

The free theory (\ref{Maxwell}) in fact admits a bigger group of $SO(2)$ duality transformations that act as ordinary rotations in the $b$-plane on the gauge potentials $A^b$ and $R^b$. If one rotates the potentials in the $b$-plane by an angle $\alpha$, the induced transformation of $(\lambda_1,\lambda_2)$ is a rotation by an angle $2\alpha$. Thus, only the combination $\lambda_1^2+\lambda_2^2$ is invariant.
If one wants to maintain the $SO(2)$ duality invariance in the interacting theory (\ref{NLED}), one must choose $g$ to depend only on this combination:
\beq
    {\cal L}={\cal L}_{Maxwell}+h(w) ,\qquad w=\sqrt{\lambda_1^2+\lambda_2^2}.\label{GDST}
\eeq
One can alternatively express $w$ as
\beq
    w=\sqrt{ - \det{\cal W}}\,,\label{w}
\eeq
with a two-dimensional `metric' ${\cal W}$ with Lorentzian signature, from comparing \eqref{GDST} and \eqref{w},
\beq
{\cal W}^{ab}\equiv  (\s H^a_{\mu\nu}-\epsilon^{ac} H^c_{\mu\nu})(\s H^b{}^{\mu\nu}-\epsilon^{bd} H^d{}^{\mu\nu})/2\,.
\eeq

In a series of recent works \cite{BLST1,Kosyakov,Bandos:2020hgy}, special attention has been given to conformal invariance in nonlinear electrodynamics,
which is expressed for theories of the form (\ref{NLED}) as
\beq
    U^{ab} f^U_{ab}+V^{ab}f^V_{ab}=f\,.
\eeq
This implies $g=\lambda_1\, \tilde g(\lambda_1/\lambda_2)$ with an arbitrary $\tilde g(x)$.
If both the $SO(2)$ symmetry and conformal invariance are imposed on the interacting theory (\ref{NLED}), only a one-parameter family of theories is left:
\beq
    {\cal L}=
    H^b\wedge \star H^b-a\,\epsilon_{bc}F^b\wedge Q^c+\delta\, w\,,
\label{Lagconf}
\eeq
where $\delta$ is a real number. 
As we shall immediately proceed to show, this is just a democratic formulation of the ModMax theory whose single-field form was introduced in \cite{BLST1}.

For the $SO(2)$-invariant theories of the form \eqref{GDST}, one can take a few further steps in understanding the conversion procedure between
the single-field and democratic formalisms, given in general by equations (\ref{galphabeta}-\ref{Lambdas}). When $g(\lambda_1,\lambda_2)=h(w)$,
as per (\ref{GDST}), one can rewrite (\ref{galphabeta}) as
\begin{equation}
\frac{\lambda_1}{w}h'=\frac{2\,\alpha}{\alpha^2+(\beta+1)^2}\,,\quad \frac{\lambda_2}{w}h'=\frac{\alpha^2+\beta^2-1}{\alpha^2+(\beta+1)^2}\,,\label{gabso2}
\end{equation}
Taking the ratio of the two equations (\ref{gabso2}), one obtains $\lambda_1(\alpha^2+\beta^2-1)=2\alpha\lambda_2$, and then, substituting (\ref{Lambdas}),
\begin{equation}
\beta^2+\frac{2s}p\alpha\beta-\alpha^2=1.
\label{SO2symm}
\end{equation}
With the expressions (\ref{lagdev}) for $\al$ and $\be$ in terms of the Lagrangian, this equation is recognized as the general $SO(2)$-invariance condition within the single-field formalism \cite{Gaillard:1981rj,Bialynicki-Birula:1984daz,Gibbons:1995cv,Hatsuda:1999ys,Bekaert:2001wa,Ivanov:2003uj,Kosyakov,Bandos:2020hgy,Svazas:2021ltf}, which can be rewritten in a simpler form using the variables $u= s+\sqrt{s^2+p^2}\,,\; v=-s+\sqrt{s^2+p^2}$:
\begin{align}
    \partial_{u}{\cal L}\,\partial_{v}{\cal L}=-1\,.\label{4dDSL}
\end{align}
We emphasize that, in the conventional single-potential formalism, this $SO(2)$ duality invariance condition is a nonlinear PDE restricting the dependence of the Lagrangian on the field strength invariants, with few known analytic solutions.
By contrast, the condition of $SO(2)$ invariance in the democratic formalism (\ref{NLED}) is a simple statement that the function $g(\lambda_1,\lambda_2)$ in the Lagrangian should only depend on the combination $\lambda_1^2+\lambda_2^2$. Given any function of one variable $h(w)$, an $SO(2)$ invariant theory is straightforwardly given by \eqref{GDST}.

To complete the analysis of the conversion between the democratic and single-field formalism in the $SO(2)$-invariant case, with (\ref{SO2symm}) and (\ref{Lambdas}), $\lambda_1=2\al\, s+2(1+\be)\,p$, and hence from (\ref{gabso2}) and (\ref{walphabetasp}),
\begin{equation}
    (\alpha s+(\beta+1)p)\,\,h'\Big|_{w=\sqrt{s^2+p^2}(\alpha^2+(\beta+1)^2)}=\alpha\sqrt{s^2+p^2}.
\label{eqSO2albe}
\end{equation}
Instead of the PDEs we obtained for the function $g$ while performing the conversion from the single-field formalism to the democratic one in the case
of general electrodynamics, with the $SO(2)$ duality invariance present, we have the above ODE for the function $h(w)$.

Finally, for the conformally invariant Lagrangian (\ref{Lagconf}), $h'=\de$, and hence (\ref{eqSO2albe}), together with (\ref{SO2symm}), is explicitly solved by
\begin{align}
 & \alpha(s,p)=-\sinh{\gamma} \frac{p}{\sqrt{p^2 + s^2}}\,,\\
 &\beta(s,p)=\sinh{\gamma}\frac{s}{\sqrt{p^2 + s^2}} - \cosh{\gamma}\,,
\end{align}
where $\delta=\coth{\frac{\gamma}{2}}$.
Using (\ref{lagdev}), these functions are integrated to the Lagrangian
\beq
    L(s,p)=-\cosh{\gamma}\,s+\sinh{\gamma}\sqrt{s^2+p^2}\,,\label{ModMax}
\eeq
which is just the Lagrangian of the ModMax theory of \cite{BLST1}.
We have thus proved that (\ref{Lagconf}) provides a democratic formulation of the ModMax theory.

\section{Democratic formulation for Abelian interactions of $p$-forms}\label{democp}

The purpose of this section is to present a generalization of what was done in the previous section for 4-dimensional nonlinear electrodynamics to arbitrary $p$-forms in $d$-dimensions. Namely, we want
to obtain a Lagrangian interacting theory of $p$-forms that explicitly features both the $p$-form gauge potential and its $(d-p-2)$-form magnetic dual. Readers specifically interested in chiral forms can proceed directly to the next section, where a similar and simpler formalism will be presented, with twice fewer gauge fields, and the derivation will look more transparent.

An important lesson learned from the previous section is that we do not have to retrace all the steps
taken under (\ref{NLE}) to construct the interaction terms. Namely, it turns out that a linear combination of the four form-fields in the theory exists that is by itself invariant under all gauge symmetries. The interaction term, to be added to the free theory Lagrangian, is then simply an arbitrary scalar function of this specific combination of gauge fields.
This structure will be used for constructing interactions that respect (twisted) selfduality both in this and the next section.

\subsection{Free theory}

As before, we start by reviewing the democratic formulation for a free $p$-form in $d$ dimensions developed in \cite{polynom}. This formulation features a $p$-form gauge potential $A_1$ and its auxiliary $p$-form partner $R_1$, as well as a $(d-p-2)$-form gauge potential $A_2$ (that will become on-shell the magnetic dual of $A_1$) and its $(d-p-2)$ form auxiliary partner $R_2$. The four corresponding field strengths are
\beq
F_1=dA_1,\qquad Q_1=dR_1,\qquad F_2=dA_2,\qquad Q_2=dR_2.
\eeq
There is additionally an auxiliary scalar $a$ that plays the same role as in the previous section (and is analogous to the auxiliary scalar of the PST theory \cite{PST1,PST2,PST3}). We remind the reader the following elementary differential form relations that play a significant role in the subsequent derivations and hold for any $(p+1)$-form $G_1$ and any $(d-p-1)$-form $G_2$: 
\beq
\s\!\s G_1= (-1)^{p+d+pd}G_1,\qquad G_1\w G_2=-(-1)^{p+d+pd}G_2\w G_1.
\eeq
The free-field democratic Lagrangian proposed in \cite{polynom} reads
\beq\label{democ}
    \mathcal{L}_{\mathrm{\,free\,democ.}}=(F_1+aQ_1)^2+(F_2+a\,Q_2)^2-2\,a\,Q_2\w F_1+2\,a\,F_2\w Q_1. 
\eeq
 (Throughout, for any form $G$, we use the notation $G^2\equiv G\w\s G=\la G, G\ra$.)
It is convenient to introduce
\beq
H_1=F_1+aQ_1\qquad\mbox{and}\qquad H_2=F_2+aQ_2,
\label{H1H2def}
\eeq
so that, up to total derivatives, (\ref{democ}) can be equivalently recast as
\beq\label{democH}
    \mathcal{L}_{\mathrm{\,free\,democ.}}=H_1^2+H_2^2+2\,da\w R_2\w H_1-2\,H_2\w da\w R_1. 
\eeq
This form is practically convenient for establishing the following set of gauge symmetries:
\begin{align}
    \de a&=0\,, \,\, \de A_1= dU\,, \,\, \de A_2=0\,,\,\, \de R_1=0\,,\,\, \de R_2=0\,; \label{maxA}\\
    \de a&=0\,, \,\, \de A_1=0\,, \,\, \de A_2=dV\,,\,\, \de R_1=0\,,\,\, \de R_2=0\,; \label{maxB}\\
    \de a&=0\,, \,\, \de A_1=0 , \,\, \de A_2=0\,,\,\, \de R_1=dU\,,\,\, \de R_2=0\,; \label{maxR}\\
    \de a&=0\,, \,\, \de A_1=0 , \,\, \de A_2=0\,,\,\, \de R_1=0\,,\,\, \de R_2=dV\,; \label{maxS}\\
    \de a&=0\,, \,\, \de A_1=-\,a\,da\w U\,, \,\, \de A_2=0\,,  \,\, \de R_1=da\w U\,,\,\, \de R_2=0\,; \label{daR}\\
    \de a&=0\,,\,\, \de A_1=0\,,\,\,\de A_2=-\,a\,da\w V\,,\,\,\de R_1=0\,,\,\, \de R_2=da\w V\,; \label{daS} \\
    \de a&=\varphi\,, \,\, \de A_1=-\,\frac{a\,\varphi}{(\del a)^2}\,i_{da}(Q_1+\s Q_2),\,\,\, \de A_2=-\,\frac{a\,\varphi}{(\del a)^2}\,\i_{da}(Q_2+(-1)^{p+d+pd}\s Q_1)\,, \nonumber\\
    &\hspace{1cm}\de R_1=\frac{\varphi}{(\del a)^2}\,\i_{da}(Q_1+\s Q_2)\,,\,\,\,
    \de R_2=\frac{\varphi}{(\del a)^2}\,\i_{da}(Q_2+(-1)^{p+d+pd}\s Q_1)\,. \label{PST3PQ}
\end{align}
The gauge parameters above are a $(p-1)$-form $U$ and $(d-p-3)$-form $V$ (specified independently in the different transformations where they appear) and a scalar $\varphi$. The first four symmetries are self-evident, as the Lagrangian only depends on the field strengths $F_{1,2}$ and $Q_{1,2}$. The next two symmetries, (\ref{daR}) and (\ref{daS}), leave $H_1$ and $H_2$ invariant and change $R_1$ and $R_2$ by something involving $da\w$. As a result, the invariance of the Lagrangian is manifest when it is written in the form (\ref{democH}). Verification of the last symmetry (\ref{PST3PQ}) is slightly more laborious. We first write
\begin{align}
&\de H_1=\varphi Q_1-\frac{\varphi}{(\del a)^2}da\w i_{da}(Q_1+\s Q_2)=\frac{\varphi}{(\del a)^2}\left(i_{da}(da\w Q_1)-da\w i_{da}\s Q_2\right),\label{Hvar}\\
&\de H_2=\varphi Q_2-\frac{\varphi}{(\del a)^2}da\w i_{da}(Q_2+(-1)^{p+d+pd}\s Q_1)=\frac{\varphi}{(\del a)^2}\left(i_{da}(da\w Q_2)-(-1)^{p+d+pd}da\w i_{da}\s Q_1\right).\nonumber
\end{align}
An important observation that will come to play a significant role later on is that
\beq
\de H_1+ \s\de H_2=0,
\label{deH1H2}
\eeq
where we have used the identities $\s\i_v (v\w A)=v\w \i_v \s A$ and  $\s (v\w \i_v A)=\i_v(v\w \s A)$ valid for any $A$ and any 1-form $v$. (A summary of differential form identities useful for our present purposes can be retrieved from the appendices of \cite{polynom}.)
The variation of the Lagrangian (\ref{democH}) under (\ref{PST3PQ}) can be written as
\begin{align}
\frac12 \de{\cal L}=H_1\w \s \de H_1+ H_2\w\s\de H_2&+ d\varphi\w R_2\w H_1+da\w \de R_2\w H_1+da\w R_2\w \de H_1\nonumber\\
& -\de H_2\w da\w R_1- H_2\w d\varphi\w R_1-H_2\w da\w \de R_1.\label{varlag}
\end{align}
We have 
\begin{align*}
&H_1\w \s \de H_1+ da\w \de R_2\w H_1=\frac{\varphi}{(\del a)^2}H_1\w \big[da\w i_{da}\s Q_1-(-1)^{p+d+pd} i_{da}(da\w Q_2)\\
&\hspace{5cm}-(-1)^{p+d+pd}da\w\i_{da}(Q_2+(-1)^{p+d+pd}\s Q_1)\big]=\varphi\, Q_2\w H_1,\\
& H_2\w\s\de H_2-H_2\w da\w \de R_1=\frac{\varphi}{(\del a)^2}H_2\w\big[da\w i_{da}\s Q_2-i_{da}(da\w Q_1)-da\w \i_{da}(Q_1+\s Q_2)\big]\\
&\hspace{8cm}=-\varphi H_2\w Q_1,\\
&da\w R_2\w \de H_1=\frac{\varphi}{(\del a)^2}da\w R_2\w\left(i_{da}(da\w Q_1)-da\w i_{da}\s Q_2\right)=\varphi \,da\w R_2\w Q_1,\\
&\de H_2\w da\w R_1=\frac{\varphi}{(\del a)^2}\left(i_{da}(da\w Q_2)-(-1)^{p+d+pd}da\w i_{da}\s Q_1\right)\w da\w R_1=\varphi\, Q_2\w da\w R_1.
\end{align*}
Putting everything together, we obtain for (\ref{varlag})
$$
\frac12 \de{\cal L}=\varphi\, dR_2\w H_1-\varphi H_2\w dR_1+\varphi \,da\w R_2\w Q_1-\varphi\, Q_2\w da\w R_1+d\varphi\w R_2\w H_1- H_2\w d\varphi\w R_1.
$$
Keeping in mind that $dH_{1}=da\w Q_{1}$ and $dH_{2}=da\w Q_{2}$, this expression is recognized as
$$
\frac12 \de{\cal L}=d\left[\varphi R_2\w H_1-(-1)^{d-p-1}\varphi\, H_2\w R_1\right].
$$
Thus, the variation of the Lagrangian under (\ref{PST3PQ}) is a total derivative, and (\ref{PST3PQ}) is a valid symmetry of (\ref{democ}).

Proceeding to the equations of motion for (\ref{democ}), we get
\begin{align}
&d[\s H_1+(-1)^{p+d+pd}aQ_2]=0,\label{eomA1}\\
&d[a\s H_1-(-1)^{p+d+pd}aF_2]=0,\label{eomR1}\\
&d[\s H_2+aQ_1]=0,\label{eomA2}\\
&d[a\s H_2-aF_1]=0,\label{eomR2}\\
&Q_1\w\s H_1+Q_2\w\s H_2-Q_2\w F_1+F_2\w Q_1=0.\label{eoma}
\end{align}
Multiplying (\ref{eomA1}) with $a$ and subtracting it from (\ref{eomR1}), we get
\beq
da\w [\s H_1-(-1)^{p+d+pd} H_2]=0.
\label{diffA1R1}
\eeq
Multiplying (\ref{eomA2}) with $a$ and subtracting it from (\ref{eomR2}), we get
\beq
da\w [\s H_2 - H_1]=0.
\label{diffA2R2}
\eeq
The Hodge dual of (\ref{diffA1R1}) is
\beq
\i_{da}[*H_2-H_1]=0.
\label{hodgediffA1R1}
\eeq
Acting with $da\w$ on (\ref{hodgediffA1R1}), acting with $\i_{da}$ on (\ref{diffA2R2}), and adding up the results, one gets
\beq
H_1=\s H_2.
\label{H1H2dual}
\eeq
This relation holds whenever the equation of motion for the gauge forms (\ref{eomA1}-\ref{eomA2}) are satisfied, and one can check that it makes the remaining equation of motion (\ref{eoma}) for the auxiliary scalar $a$ automatically satisfied.

Substituting (\ref{H1H2dual}) back into (\ref{eomA1}) and (\ref{eomA2}), we get
\beq
da\w dR_2=0,\qquad da\w dR_1=0.
\eeq
These equations are integrated in full generality as
\beq
R_1=dB_1+da\w C_1,\qquad R_2=dB_2+da\w C_2,
\eeq
where $B_1$ and $C_1$ are arbitrary $(p-1)$-forms, and $B_2$ and $C_2$ are arbitrary $(d-p-3)$-forms. (A description of the integration procedure can be found in Appendix C of \cite{polynom}.) Furthermore, the above expressions can always be gauge-transformed to zero using (\ref{maxR}-\ref{daS}), yielding
\beq
R_1=0,\qquad R_2=0.
\label{R1R2zero}
\eeq
Thus $R_1$ and $R_2$ (as well as the auxiliary scalar $a$) are pure gauge degrees of freedom. Furthermore, substituting (\ref{R1R2zero}) into (\ref{H1H2dual}), we get
\beq
F_1=\s F_2,
\label{F1F2dual}
\eeq
which says precisely that $F_1$ is 
dual to $F_2$ while both $A_1$ and $A_2$ satisfy the free equations of motion $d\s F_1=0$, $d\s F_2=0$.


\subsection{Interactions}

In section~\ref{nled}, we reviewed a particular approach, introduced in \cite{democ}, to adding interactions on top of free-field democratic Lagrangians of the sort given by (\ref{democ}). The essence of this approach, in the language of this section, is that $H_1$ and $H_2$ defined by (\ref{H1H2def}) are by themselves invariant under the gauge theory transformations of the free theory (\ref{maxA}-\ref{daS}) but not under the last remaining gauge
transformation (\ref{PST3PQ}). Thus, one may try to add to the free Lagrangian an arbitrary scalar function of $H_1$ and $H_2$ and then attempt
constraining this function in such a way that the last crucial gauge symmetry (\ref{PST3PQ}) emerges.

The above procedure is precisely what has been successfully implemented in \cite{democ} and reviewed in section~\ref{nled}. In retrospect, however,
one may do wiser than that. Namely, from (\ref{deH1H2}), $H_1+ \s H_2$ is invariant under {\it all} the gauge transformations (\ref{maxA}-\ref{PST3PQ}) of the free theory \cite{democ}. Thus, a convenient way to construct a very large class of interacting theories that automatically respect the same gauge symmetries as the free Lagrangian is to add to (\ref{democ}) an arbitrary scalar function $\cal F$ of $H_1+ \s H_2$,
\beq\label{democX}
    \mathcal{L}_{\mathrm{\,democ.}}=(F_1+aQ_1)^2+(F_2+a\,Q_2)^2-2\,a\,Q_2\w F_1+2\,a\,F_2\w Q_1 +\mathcal{F}(H_1+\s H_2). 
\eeq
This structure was originally guessed by examining the output of the considerations reviewed in section~\ref{nled}, and briefly summarized in the conclusions of \cite{democ}. We shall now explain in more detail how it works.

With the inclusion of the new term in (\ref{democX}), the equtions of motion (\ref{eomA1}-\ref{eoma}) are modified as
\begin{align}
&d[\s H_1+(-1)^{p+d+pd}aQ_2+\s X]=0,\label{eomA1X}\\
&d[a\s H_1-(-1)^{p+d+pd}aF_2+a\s X]=0,\label{eomR1X}\\
&d[\s H_2+aQ_1-X]=0,\label{eomA2X}\\
&d[a\s H_2-aF_1-aX]=0,\label{eomR2X}\\
&Q_1\w\s H_1+Q_2\w\s H_2-Q_2\w F_1+F_2\w Q_1+aQ_1\w \s X-aQ_2\w X=0.\label{eomaX}
\end{align}
We have introduced $X$ to denote the following $(p+1)$-form-valued function\footnote{Note that the differentiation is somewhat subtle since the components of the form field are not independent, and the result depends on whether one identifies the related components before or after differentiation. The ambiguity, however, is a pure numerical factor that can be absorbed, if desired, into a redefinition of $\cal F$. We ignore such inconsequential factors, here and in similar formulas below, so as not to clutter the formulas.} of $H_1+\s H_2$ obtained by differentiating $\cal F$:
\beq
X_{i_1\cdots i_{p+1}}\equiv \frac{\del\mathcal{F}(Y)}{\del Y^{i_1\cdots i_{p+1}}}\Bigg|_{Y=H_1+\s H_2}.
\label{defXp}
\eeq

The treatment of these equations of motion is directly analogous to the free case. First, multiplying (\ref{eomA1X}) with $a$ and subtracting it from (\ref{eomR1X}) yields
\beq
da\w[\s H_1-(-1)^{p+d+pd} H_2+\s X]=0.
\label{daH1X}
\eeq
Then, multiplying (\ref{eomA2X}) with $a$ and subtracting it from (\ref{eomR2X}) yields
\beq
da\w[\s H_2-H_1-X]=0.
\label{daH2X}
\eeq
The Hodge dual of (\ref{daH1X}) is
\beq
\i_{da}[\s H_2-H_1-X]=0.
\eeq
Acting on this with $da\w$, acting with $\i_{da}$ on (\ref{daH2X}), and adding the results yields
\beq
H_1+X=\s H_2.
\label{H1XH2}
\eeq
Substituting this equation back into (\ref{eomA1X}) and (\ref{eomA2X}) yields
\beq
da\w dR_2=0,\qquad da\w dR_1=0,
\eeq
identical to the free case, integrated in full generality by $R_1=dB_1+da\w C_1$ and $R_2=dB_2+da\w C_2$, then gauge-transformed to
\beq
R_1=0,\qquad R_2=0,\qquad H_1=F_1,\qquad H_2=F_2,
\eeq
so that the only remaining propagating fields are $A_1$ and $A_2$ related to each other by the nonlinear twisted selfduality relation
\beq
F_1-\s F_2+X=0
\label{F1XF2}
\eeq
that follows immediately from (\ref{H1XH2}). After $R_1$ and $R_2$ have been gauge-transformed to zero, one should understand $X$ as a function of $F_1+*F_2$ instead of $H_1+*H_2$ as in (\ref{defXp}). Note that the nonlinear selfduality relation (\ref{F1XF2}) deforms the free selfduality relation $F_1-\s F_2=0$ by an arbitrary function of the opposite chirality combination $F_1+\s F_2$, which is the same structure (with the opposite sign conventions) that emerged in (\ref{FF}) from the four-dimensional derivation, as anticipated by the observations in the conclusions of \cite{democ} and confirmed in detail by our present derivations. Another observation is that the same gauge invariant combination $H_1+\s H_2$ can be used to describe interactions with other fields. In particular, supplementing the free Lagrangian \eqref{democH} with a term $(H_1+\s H_2)\wedge \s T$, where $T$ is a $(p+1)$-form constructed from other fields in the theory, the corresponding equation \eqref{F1XF2} will describe the deformation of the free (Abelian) $p$-form theory including interactions with other fields.

\section{Abelian interactions of chiral $2k$-forms in $4k+2$ dimensions}\label{chiralf}

When the rank of the field strength is one half of the number of spacetime dimensions, that is,
for a $p$-form gauge potential $A$ in $d=2p+2$ dimensions, the magnetic dual of $A$ is also a $p$-form. This allows one to transform the two fields into each other by duality rotations, discussed for electrodynamics in Section \ref{nled}.
Furthermore, when $p=2k$ and hence $d=4k+2$, one can identify $A$ and its magnetic dual, leaving a single field known as a chiral (or selfdual) form. In this section, we shall discuss the Lagrangian description of such chiral forms, starting with the free case, and then proceeding to include interactions.

\subsection{Free theory}

For the free case, we consider the following Lagrangian, first proposed in \cite{Mkrtchyan:2019opf}:
\beq\label{FQlag}
    \mathcal{L}_{\mathrm{\,free\,chiral}}=(F+a\,Q)^2+2\,aF\w Q\,,
\eeq
where $F\equiv dA$ is the $(2k+1)$-form field strength of the $2k$-form field $A$, $Q\equiv dR$ is the 
$(2k+1)$-form field strength of the $2k$-form field $R$, and $a$ is an auxiliary scalar field. (Throughout, for any form $G$, we use the notation $G^2\equiv G\w\s G$.)
It may appear at first sight that we are dealing with a large set of dynamical fields (while our aim is to describe
a single chiral $2k$-form). The truth is that most of the Lagrangian fields will be eliminated on-shell, leaving only the desired chiral form, which happens due to the following large set of gauge symmetries:
\begin{align}
    \de a&=0\,, \,\, \de A= dU\,, \,\, \de R=0\,; \label{Asymm} \\
    \de a&=0\,, \,\, \de A=0 , \,\, \de R=dU\,; \label{Rsymm} \\
    \de a&=0, \,\, \de A=-\,a\,da\w U\,,  \,\, \de R=da\w U\,; \label{ARgauge}\\
\de a&=\varphi\,, \,\, \de A=-\,\frac{a\,\varphi}{(\del a)^2}\,\i_{da}(Q+\s Q)\,, \,\, \de R=\frac{\varphi}{(\del a)^2}\,\i_{da}(Q+\s Q)\,.\label{PST3FQ}
\end{align}
The parameters in these transformations are an arbitrary position-dependent scalar $\varphi$ and an arbitrary position-dependent $(p-1)$-form $U$ (the three copies of $U$ appearing in the first three transformations designate independent parameters). 

The first two gauge symmetries (\ref{Asymm}-\ref{Rsymm}) are ordinary gradient shifts of gauge fields,
and are obvious since (\ref{FQlag}) only depends on the field strength $F$ and $Q$, but not on the potentials $A$ and $R$. To manifest the remaining two symmetries, it is convenient to first introduce
\beq
H\equiv F+aQ\,,
\label{defH}
\eeq
and rewrite (\ref{FQlag}), up to total derivatives, as
\beq
\mathcal{L}_{\mathrm{\,free\,chiral}}=H^2+2da\w H\w R\,.
\label{FQlagH}
\eeq
The invariance under (\ref{ARgauge}) is now manifest, since $H$ is invariant by itself, while $R$ transforms by something involving $da\w$, and $da\w da=0$. To see the invariance under (\ref{PST3FQ}), we write
\beq
\de H=\varphi Q - \frac{\varphi}{(\del a)^2}da\w \i_{da}(Q+\s Q)=\frac{\varphi}{(\del a)^2}\left[\i_{da}(da\w Q)-da\w\i_{da}\s Q\right]\,.\label{Hvarchi}
\eeq
In particular, $\de H$ satisfies
\beq
\de H+\s \de H=0\,,
\label{deHsH}
\eeq
which will come to play an important role when deforming (\ref{FQlag}) to include interactions.
It is useful to keep in mind, here and for the subsequent derivations, that for any $(2k+1)$-forms $G$ and $\tilde G$ in $4k+2$ spacetime dimensions
\beq
\s\!\s G= G,\qquad G\w\tilde G=-\tilde G\w G\,,
\eeq
and additionally, that for any 1-form $v$ and any $A$,
\beq
\s (v\w i_v A)=i_v(v\w \s A),\qquad \s \i_v (v\w A)=v\w i_v\s A,\qquad v\w\i_v A+\i_v(v\w A)=v^2 A\,.
\label{vAid}
\eeq
A summary of useful differential form identities can be located in the appendices of \cite{polynom}.
The variation of (\ref{FQlagH}) under \eqref{PST3FQ} is
\beq
\frac12 \de\mathcal{L}=H\w\s\de H+d\varphi\w H\w R+da\w\de H\w  R+ da\w H\w\de R\,.
\label{varlagchi}
\eeq
Then,
\begin{align*}
&H\w\s\de H+ da\w H\w\de R=\frac{\varphi}{(\del a)^2}H\w\left[da\w\i_{da}\s Q-\i_{da}(da\w Q)-da\w\i_{da}(Q+\s Q)\right]=-\varphi H\w Q\,,\\
&da\w\de H\w  R=\frac{\varphi}{(\del a)^2}da\w\left[\i_{da}(da\w Q)-da\w\i_{da}\s Q\right]\w R=
\varphi\, da\w Q\w R\,,
\end{align*}
where the last line uses $da\w da=0$ and $v\w i_v(v\w A)=v\w[i_v(v\w A)+v\w i_v A]=v^2\,v\w A$.
Plugging these expressions back into (\ref{varlagchi}), we get
\beq
\frac12 \de\mathcal{L}=-\varphi H\w Q+\varphi\, da\w Q\w R+d\varphi\w H\w R=d(\varphi H\w R),
\eeq
where we have used $dH=da\w Q$. Thus, the variation of the Lagrangian under (\ref{PST3FQ}) is a total derivative, and the symmetry (\ref{PST3FQ}) is respected by (\ref{FQlag}).

We then turn to the equations of motion given by
\begin{align}
&d[\s H+aQ]=0,\label{eomchiA}\\
&d[a\s H-aF]=0,\label{eomchiR}\\
&Q\w\s H+F\w Q=0.\label{eomchia}
\end{align}
Multiplying (\ref{eomchiA}) by $a$ and subtracting it from (\ref{eomchiR}) yields
\beq
da\w(\s H-H)=0,
\label{dasHH}
\eeq
the Hodge dual of which is
\beq
\i_{da}(\s H-H)=0.
\label{idasHH}
\eeq
Acting with $\i_{da}$ on (\ref{dasHH}) and with $da\w$ on (\ref{idasHH}) and adding up the results, by (\ref{vAid}),
\beq
H=\s H.
\label{HsH}
\eeq
This relation is satisfied whenever the form-field equations of motion (\ref{eomchiA}-\ref{eomchiR}) are satisfied, and it automatically implies the equation of motion (\ref{eomchia}) for the auxiliary scalar $a$ (remembering that $F\w Q=-Q\w F$ and $Q\w Q=0$).

Plugging (\ref{HsH}) back into (\ref{eomchiA}) yields
\beq
da\w dR=0,
\eeq
which is solved in full generality (see Appendix C of \cite{polynom} for a more detailed explanation) by
\beq
R=dB+da\w dC,
\eeq
with $B$ and $C$ being arbitrary $(2k-1)$-forms. Then, $R$ can always be gauge-transformed to zero using (\ref{Rsymm}-\ref{ARgauge}), so that
\beq
R=0,\qquad H=F.
\eeq
Therefore, (\ref{HsH}) implies
\beq
F=\s F,
\label{FsF}
\eeq
which is exactly the desired free-field selfduality relation. Thus, $R$ has been gauged away on-shell, and $a$ is a pure gauge degree of freedom that can be arbitrarily shifted by (\ref{PST3FQ}), leaving the chiral $2k$ form $A$ satisfying (\ref{FsF}) as the only physical degree of freedom, as intended.


\subsection{Interactions}

Our inclusion of interactions into (\ref{FQlag}) is guided by preserving its symmetries (\ref{Asymm}-\ref{PST3FQ}). There are two crucial observations in this regard. First, that $H$ defined by (\ref{defH}) is by itself invariant under (\ref{Asymm}-\ref{ARgauge}), but not under (\ref{PST3FQ}).
Second, that due to (\ref{deHsH}), $H+\s H$ is invariant under all the symmetries (\ref{Asymm}-\ref{PST3FQ}) of the free Lagrangian (\ref{FQlag}).
As a result, adding an arbitrary scalar function $\cal F$ of $H+*H$ to the free Lagrangian (\ref{FQlag}) automatically produces an interacting theory that respects the symmetries (\ref{Asymm}-\ref{PST3FQ}). We shall then proceed to explore the equations of motion of the resulting Lagrangian
\beq
\mathcal{L}_{\mathrm{\,chiral}}=H^2+2aF\w Q +\mathcal{F}(H+*H),\qquad H\equiv F+aQ,
\label{lagnl}
\eeq
given by
\begin{align}
&d[\s H+aQ+\s X-X]=0,\label{eomchiAX}\\
&d[a\s H-aF+a\s X-aX]=0,\label{eomchiRX}\\
&Q\w\s H+F\w Q+Q\w(\s X-X)=0.\label{eomchiaX}
\end{align}
Here,
\beq
X_{i_1\cdots i_{2k+1}}\equiv \frac{\del\mathcal{F}(Y)}{\del Y^{i_1\cdots i_{2k+1}}}\Bigg|_{Y=H+\s H}
\label{defX}
\eeq
is a rank $2k+1$ fully antisymmetric tensor -- that is, a $(2k+1)$-form -- obtained by differentiating the scalar function $\cal F$ with respect to the components of its argument. 

As in the preceding free theory derivations, multiplying (\ref{eomchiAX}) with $a$ and subtracting it from (\ref{eomchiRX}) yields
\beq
da\w(\s H-H+\s X-X)=0.
\eeq
By an argument identical to the one displayed under (\ref{dasHH}), this implies that
\beq
\s\! H-H+\s X-X=0,
\label{sHX}
\eeq
and hence the equation of motion for $a$ given by (\ref{eomchiaX}) is automatically satisfied. Furthermore, plugging (\ref{sHX}) into (\ref{eomchiAX}) yields
\beq
da\w dR=0,
\eeq
exactly the same equation as in the free theory, which is again integrated as $R=dB+da\w dC$, which can be gauged to $R=0$ and hence $H=F$. Thereafter, from (\ref{sHX}),
\beq
\s\! F-F+\s X-X=0,
\label{FFXX}
\eeq
where $F$ should now be substituted instead of $H$ in the definition of $X$, in other words, 
\beq
X_{i_1\cdots i_{2k+1}}\equiv \frac{\del\mathcal{F}(Y)}{\del Y^{i_1\cdots i_{2k+1}}}\Bigg|_{Y=F+\s F}.
\label{defXF}
\eeq

To summarize, a very simple and general nonlinear selfduality relation (\ref{FFXX}) resulted as the equation of motion for the only propagating
degree of freedom of the Lagrangian (\ref{lagnl}). The nonlinearities are contained in the $(2k+1)$-form valued function $X(F+*F)$ obtained by differentiating an arbitrary scalar function $\cal F$ of a selfdual $(2k+1)$-form $Y$. 
Interactions with other fields can be also included into $\cal F$, which may have dependence on other fields. In particular, the interaction term ${\cal F}= (H+\s H)\wedge T$ where $T$ is a $(2k+1)$-form sourced by any other fields in the theory, gives rise to the same equation \eqref{FFXX} where now self-interactions are replaced by interactions with additional fields.

As $k$ increases in $4k+2$, the relevant spacetime dimensions that are most commonly discussed in physics literature are 2, 6 and 10. In 2 dimensions, we are dealing with a scalar whose `field strength' is a vector. Since only one functionally independent scalar can be constructed from a vector, and this scalar vanishes for a self-dual vector in two spacetime dimensions, no interaction terms described by $\cal F$ in (\ref{lagnl}) exist in this case. This is in accord with the perception that chiral scalars in 2d cannot self-interact. Nontrivial interactions arise starting from the much-discussed case of dimension 6. We shall proceed to analyze this case in more detail, followed by the case of chiral 4-forms in 10 dimensions.


\subsection{Chiral 2-forms in 6 dimensions}\label{chi6d}

The problem of classifying theories of the form (\ref{lagnl}) is the problem of classifying all functionally independent scalars made of a selfdual form $Y\equiv H+\s H$. Then, in a general theory of the form (\ref{lagnl}), $\cal F$ can be thought of as an arbitrary function of these scalars.

In six spacetime dimensions, there exists only one functionally independent scalar one may construct from a selfdual form $Y$, that is, a fully antisymmetric rank 3 tensor satisfying
\beq
Y_{ijk}=\frac16 \eps_{ijklmn}Y^{lmn}.
\label{sddef}
\eeq
This invariant can be chosen as
\beq
I_4^{\mathrm{(6d)}}\equiv M_i^{\,\,j}M^{\,\,i}_j=\Tr[M^2],
\label{defI4}
\eeq
where
\beq
M_i^{\,\,j}\equiv Y_{ikl}Y^{jkl}.
\label{Mdef}
\eeq
It may appear surprising at first that only one independent invariant exists, but the surprise is perhaps mitigated by the observation that a selfdual 3-form in 6 dimensions has $\frac{6!}{2(3!)^2}=10$ components, while the number of independent Lorentz transformations $6\cdot 5/2=15$ is bigger than that, leaving little room for invariant information. 

We demonstrate how any polynomial invariant made of $Y$ can be re-expressed through (\ref{defI4}) in Appendix \ref{6d invariant}.
In addition to this hands-on argument that applies chiral form identities to express any invariant of $Y$ through (\ref{defI4}), 
we mention the following shortcut that reaches the same conclusion in the language of spinor representations (see, e.g., \cite{Park}), keeping in mind 
the isomorphism between $so(1,5)$ and $su^*(4)$. In the spinor language, the selfdual three-form is a symmetric Weyl bispinor $Y_{ab}$ (or a symmetric bifundamental representation of $su^*(4)$, $a,b=1,\dots,4$). The only invariant tensor we can contract with products of $Y_{ab}$ so as to form an invariant is the Levi-Civita tensor $\epsilon^{a_1a_2a_2a_4}$. On the other hand, since $Y_{ab}$ is symmetric, only one index of it can be contracted to a given Levi-Civita tensor. Then, any Levi-Civita tensor comes with the following contraction:
\begin{align}
    \epsilon^{a_1a_2a_3a_4}Y_{a_1b_1}\,Y_{a_2b_2}\,Y_{a_3b_3}\,Y_{a_4b_4}=\frac{1}{24} J^{(6d)}_4\,\epsilon_{b_1b_2b_3b_4}\,,
\end{align}
proportional to the invariant:
\begin{align}
    J^{(6d)}_4=\epsilon^{a_1a_2a_3a_4}\epsilon^{b_1b_2b_3b_4}\,Y_{a_1b_1}\,Y_{a_2b_2}\,Y_{a_3b_3}\,Y_{a_4b_4}\,,
\end{align}
which is thus the only functionally independent invariant translating in the Lorentz tensor language to $I^{(6d)}_4$ given by \eqref{defI4}, up to a numerical factor.\footnote{Another elegant argument was provided to us by Amihay Hanany: the symmetric bispinor $Y_{\alpha\beta}$ breaks the 15-dimensional algebra $su(4)$ to its 6-dimensional subalgebra $so(4)$ leaving $Y$ invariant; the remaining $15-6=9$ generators transform the 10-dimensional representation $Y_{\alpha\beta}$ non-trivially, leaving room for only $10-9=1$ invariant.}

Since we can express any invariant of a selfdual 3-form in six spacetime dimensions through $I_4^{\mathrm{(6d)}}$, we can write the most general Lagrangian (\ref{lagnl}) in $d=6$ as
\beq
\mathcal{L}
=H^2+2aF\w Q +\mathcal{F}(I_4^{\mathrm{(6d)}}),\qquad H\equiv F+aQ,
\label{chiral6d}
\eeq
where $\cal F$ is an arbitrary function of one variable, and explicitly,
\beq
I_4^{\mathrm{(6d)}}= Y_{ikl}Y^{jkl}Y^{imn}Y_{jmn},\qquad Y=H+\s H.
\label{I4d6}
\eeq
It is straightforward to show by a simple dimensional argument that among this general class of self-interacting chiral 2-form theories, there is a one-parameter family that also respects conformal symmetry, similarly to the four-dimensional case \eqref{Lagconf}:
\begin{align}
    \mathcal{L}_{\mathrm{\,conf}}=H^2+2aF\w Q +\delta\,\sqrt{I_4^{\mathrm{(6d)}}}\,.
\label{confchiral6d}
\end{align}
This is a covariant Lagrangian for the conformal chiral 2-form electrodynamics theories of \cite{Townsend1,Bandos:2020hgy}.

We conclude with comparing the equations of motion of \eqref{chiral6d} with another formulation
of chiral 2-form interactions due to Perry and Schwarz \cite{PSch}. That formulation is given in terms of a five-dimensional reduction of the six-dimensional degrees of freedom, in line with the belief prevalent at the time that no manifestly Lorentz-covariant formulation of chiral form interaction is viable. The core of our present work is precisely to demonstrate that not only are there simple and natural Lorentz-covariant equations of motion, but they can also be derived from a local, Lorentz-invariant Lagrangian.\footnote{To provide a more complete historical perspective, we mention that a Lorentz-covariant (PST) Lagrangian formalism \cite{PST3} was developed for free chiral field concurrently with \cite{PSch}, based on earlier ideas from \cite{PST1,PST2}. It was later extended to include interactions in 6 spacetime dimensions starting with \cite{S5M}, see section 3 of \cite{Buratti:2019guq} for an accurate recent summary. In 6 dimensions, this formalism covers the same range of theories as our present construction. The true power of our formalism is revealed in higher dimensions, where only limited progress has been achieved using the PST approach \cite{Buratti:2019guq}. We thank Dmitri Sorokin for consultations on the history of covariant Lagrangian theories for chiral forms.}  Nonetheless, to compare with the construction of \cite{PSch}, we will dimensionally reduce \eqref{chiral6d} to five dimensions.

In order to perform the desired dimensional reduction, we introduce the six-dimensional unit vector $n=(0,0,0,0,0,1)$ and decompose the fields $A$ and $R$ whose field strengths $F$ and $Q$ appear in (\ref{chiral6d}) in terms of their projections along and perpendicular to this vector:
\beq
A_1\equiv \i_n A, \qquad A_2=\i_n(n\w A), \qquad R_1\equiv \i_n R, \qquad R_2=\i_n(n\w R).
\eeq
$A_1$ and $R_1$ are one-forms, and $A_2$ and $R_2$ are two-forms. By the standard projection-rejection identity
\beq
A=n\w A_1+A_2, \qquad R=n\w R_1+R_2.
\eeq
To perform the dimensional reduction to five dimensions, we simply assume that all the fields are constant along the last direction (the direction of $n$). In that case,
\beq
F=-n\w F_1+F_2, \qquad Q=-n\w Q_1+Q_2,
\eeq
with $F_1\equiv dA_1$, $F_2\equiv dA_2$, $Q_1\equiv dR_1$, $Q_2\equiv dR_2$. Given that $n$ is constant and all the fields are constant in the direction of $n$, we have $\i_n F_1=\i_n Q_1=0$, $\i_n F_2= \i_n Q_2=0$, $F_2=\i_n(n\w F_2)$, $Q_2=\i_n(n\w Q_2)$. Thus, one has
\begin{align}
&(F+aQ)^2=(F_1+aQ_1)^2+(F_2+aQ_2)^2,\\
&F\w Q=n\w (F_2\w Q_1-F_1\w Q_2),\\
&Y\equiv F+aQ+\s (F+aQ)= - n\w \tilde Y - \i_n \s\tilde Y,\qquad \tilde Y\equiv F_1+a Q_1 - \i_n\s (F_2+aQ_2).\label{YYtilde}
\end{align}
Note that $\sfi\equiv -\i_n\s$ is precisely the five-dimensional Hodge star operation with respect to the five directions orthogonal to $n$. From these relations, we observe that dimensional reduction of the six-dimensional chiral theory (\ref{lagnl}) produces precisely the five dimensional democratic theory (\ref{democX}) for a 1-form $A_1$ and its dual 2-form $A_2$. The $\cal F$-term of this theory is, most generally, an arbitrary invariant made of the 2-form $\tilde Y=H_1+\sfi H_2$. There can be only two such independent invariants (which is, in particular, the number of independent nonzero eigenvalues of a real antisymmetric $5\times 5$ matrix). Following \cite{PSch}, we can choose these invariants as
\beq
\zz_1=\frac12\, \tilde Y_{\al\be}\tilde Y^{\be\al},\qquad \zz_2=\frac14\, \tilde Y_{\al\be}\tilde Y^{\be\gamma}\tilde Y_{\gamma\de}\tilde Y^{\de\al},
\eeq
where, until the end of this section, we use Greek letters from the beginning of the alphabet to denote the five-dimensional directions. Most generally, within the context of (\ref{democX}), $\cal F$ is an arbitrary function of
$\zz_1$ and $\zz_2$. That cannot be so, however, for theories resulting from dimensionally reducing (\ref{chiral6d}), since $\cal F$ in these theories only depends on a single scalar. Substituting (\ref{YYtilde}) into the (\ref{chiral6d}) results in an $\cal F$-term depending on a single scalar made of $\tilde Y$. Since this scalar is quartic, it must be a linear combination of $\zz_1^2$ and $\zz_2$. 
The coefficients can be fixed by expanding $I_4^{\mathrm{(6d)}}$ through the components of $Y$, giving
\beq
\frac1{24}\, I_4^{\mathrm{(6d)}}=-\zz_1^2+{4}\,\zz_2\,.
\eeq

Thus, any theory of the form (\ref{chiral6d}) is expressed, after the dimensional reduction to five dimensions, as
\beq\label{democ5d}
    \mathcal{L}_{\mathrm{\,5d}}=(F_1+aQ_1)^2+(F_2+a\,Q_2)^2-2\,a\,Q_2\w F_1+2\,a\,F_2\w Q_1 +h(\zz_1^2-{4}\zz_2). 
\eeq
The corresponding equation of motion (\ref{F1XF2}) for the propagating degrees of freedom is
\beq
F_1-\sfi F_2 - \big[{2}\,\zz_1 (F_1+\sfi F_2)- {4}\, (F_1+\sfi F_2)^3\big] h'=0,
\label{F1F25}
\eeq
where $(F_1+\sfi F_2)^3$ denotes the ordinary matrix cube of the $5\times 5$ matrix $F_1+\sfi F_2$.
To compare with \cite{PSch}, we need to resolve this system of algebraic equations to express $F_1$ through $F_2$.

We emphasize that the steps we are presently taking are directly analogous to the exposition of section~\ref{singlef} where we were converting the democratic representation of 4d electrodynamics to the ordinary single-field form, except that now we are in 5d and the electric-magnetic dual potentials are a 1-form and a 2-form. Equation (\ref{F1F25}) plays a role identical to (\ref{FF}), and 
one expects that it can be resolved in a manner analogous to (\ref{F1F2}) to express $\sfi F_2$ through $F_1$
\beq
\sfi\!F_2=\alpha(y_1,y_2) F_1+\be (y_1,y_2) (F_1)^3,
\label{albeF1}
\eeq
where\footnote{Indeed, the only way to construct a covariant rank 2 tensor from a rank 2 tensor is matrix powers and multiplication by invariants. Furthermore, for an antisymmetric rank 2 tensor in 5d, all even matrix powers are symmetric tensors and have a vanishing antisymmetric part, while any matrix powers higher than 3 are expressible through the tensor itself, its matrix cube, and invariants.}
\beq
y_1\equiv \frac12\Tr[(F_1)^2],\qquad y_2\equiv \frac14\Tr[(F_1)^4]
\eeq
are the two invariants one can construct from $F_1$. This is precisely the form of the five-dimensional equations of motion in \cite{PSch}. To explore the constraints on $\alpha$ and $\beta$ imposed by (\ref{F1F25}), we substitute (\ref{albeF1}) into (\ref{F1F25}) to obtain
\beq
(1-\al)F_1-\be[F_1]^3 +\left\{(1+\al)F_1+\be[F_1]^3\right\}\left\{{2}\zz_1-{4}((1+\al)F_1+\be[F_1]^3)^2\right\}h'=0\,,
\label{F15}
\eeq
where $\zz_1=\frac12\Tr[\{(1+\al)F_1+\be[F_1]^3\}^2]$. This is an algebraic equation for the real antisymmetric matrix $F_1$. To analyze this equation, we follow the strategy of the appendix  of \cite{PSch},  and assume that the matrix $F_1$ can be Lorentz-rotated to the form where all of its components are zero, except for
\beq
(F_1)_{12}=-(F_1)_{21}=\lambda_+\,,\qquad (F_1)_{34}=-(F_1)_{43}=\lambda_-\,.
\eeq
(This is, strictly speaking, only possible if $\Tr[F_1^2] < 0$, but one may expect that the resulting analytic relations will hold in general.)
In terms of $\lambda_\pm$, 
\beq
\zz_1=-[(1+\al)\lambda_+-\be\lambda_+^3]^2-[(1+\al)\lambda_--\be\lambda_-^3]^2\,. 
\eeq
One then rewrites (\ref{F15}) equivalently as
\begin{align}
&(1-\al)\lambda_++\be\lambda_+^3 -\left\{2\,\zz_1((1+\al)\lambda_+-\be\lambda_+^3)+4\,((1+\al)\lambda_+-\be\lambda_+^3)^3\right\}h'=0\,,\label{lphp}\\
&(1-\al)\lambda_-+\be\lambda_-^3 -\left\{2\,\zz_1((1+\al)\lambda_--\be\lambda_-^3)+4\,((1+\al)\lambda_--\be\lambda_-^3)^3\right\}h'=0\,.\label{lmhp}
\end{align}
These equations can be recast as
\begin{align}
&1-\al+\be\lambda_+^2 -(1+\al-\be\lambda_+^2)\,\tilde h=0\,,\\
&1-\al+\be\lambda_-^2 +(1+\al-\be\lambda_-^2)\,\tilde h=0\,,
\end{align}
where $\tilde h\equiv  2[\{[(1+\al)\lambda_+-\be\lambda_+^3]^2-[(1+\al)\lambda_--\be\lambda_-^3]^2\}h'$.
Then, eliminating $\tilde h$ yields
\beq
 \al^2 - (\lambda_+^2+\lambda_-^2)\al\be+ \lambda_+^2\lambda_-^2\be^2 =1\,.
\eeq
Keeping in mind that $y_1= - (\lambda_+^2+\lambda_-^2)$ and $y_2=( \lambda_+^4+\lambda_-^4)/2$, this can be rewritten as
\beq
 \al^2+y_1\al\be+ \left(\frac{y_1^2}2-y_2\right)\be^2 =1,
\eeq
exactly the condition derived in \cite{PSch} from the requirement that the dimensionally reduced 5d theory uplifts to a Lorentz-invariant theory in 6d. In our derivation, this condition emerged automatically from dimensionally reducing the theory (\ref{chiral6d}) to 5d. It plays a role very similar to the $SO(2)$ duality invariance condition (\ref{SO2symm}) in 4d nonlinear electrodynamics.

We have thus not only provided simple and explicit Lorentz covariant equations of motion in 6d for all chiral form theories in the class described in \cite{PSch}, but also a Lagrangian formulation for all of these equations of motion. We will further show in section~\ref{eoms} that the most general conceivable equations for chiral forms in 6d can be reduced to this form.


\subsection{Chiral 4-forms in 10 dimensions}\label{chi10d}

If one attempts to classify all invariants of a selfdual form in 10 dimensions, there is still a large choice of identities that reduce the number of independent invariants, but it is much more challenging than in 6 dimensions to manage them in any explicit form, and the number of invariants is expected to be large. The number of independent components of a selfdual 5-form in 10 dimensions is $\frac{10!}{2(5!)^2}=126$ while the number of Lorentz generators is $10\cdot 9/2=45$. Thus, naively, one expects at least $126-45=81$ independent invariants, while obviously not more than 126.

A systematic theory of Lie group invariants exists and revolves around the Hilbert series and the Molien (or Molien-Weyl) formula \cite{MW}. It has been successfully applied\footnote{We thank Prarit Agarwal, Kirsty Gledhill, Julius Grimminger and Amihay Hanany for discussions on this matter.} to other problems in high-energy theory \cite{MW1,MW2}, and explicit classifications of invariants for the simpler case of tensors in three Euclidean dimensions exist in the literature \cite{tensinv1,tensinv2}, based on precisely these techniques. (See also \cite{Agarwal:2020wdf}.)
It seems challenging, however, to apply these techniques to our situation in $d\geq 10$,
and we will refrain from pursuing this strategy here.

A more practical question, which is much more manageable, is to classify low-order invariants. Indeed, at least in the particle theory context, one would rarely be concerned with, say, 40-particle interactions resulting from an invariant in the Lagrangian obtained from a product of 40 copies of the fundamental fields.
It is natural to focus on low-order invariants first, and we shall do it for quartic invariants of selfdual 5-forms in 10 dimensions.

It turns out that there is only one functionally independent quartic invariant of a chiral 5-form $Y$ in 10 dimensions, which can be chosen as
\beq
I_4^{\mathrm{(10d)}}= Y_{iklmn}Y^{jklmn}Y^{ipqrs}Y_{jpqrs}.
\label{I410d}
\eeq 
We summarize arguments that lead to this conclusion in Appendix~\ref{inv10d} in parallel to our proof in Appendix~\ref{6d invariant} that, in six dimensions, there is only one independent invariant (without any restrictions on the polynomial degree).
We furthermore provide here an alternative shortcut in the spinor language leading to the same conclusion. A simple observation is that symmetric Majorana-Weyl bispinors in ten dimensions parametrize a (136-dimensional) vector space isomorphic to the direct sum of the vector spaces of selfdual five-form (126 dimensions) and vector (10 dimensions) representations of the Lorentz algebra. In the language of the Clifford algebra (see \cite{Park}), this means that the ten symmetric $16\times 16$ $\gamma$-matrices $\gamma_\mu^{ab}$ can be used to construct invariant spin-tensors $\Sigma^{ab,cd}=\gamma^{ab}_\m\gamma^\m{}^{cd}$,
that can contract spinor indices of the symmetric bispinors $H_{ab}$ parametrizing the selfdual five-forms (and satisfying $\gamma_\mu^{ab}H_{ab}=0$). Then, since $\Sigma^{(ab,c)d}=0$\,, it follows immediately that $\Sigma^{ab,cd}H_{ac}=0=\Sigma^{ab,cd}H_{ab}$ and the only possible contractions are those where each $\Sigma$ can contract at most one index from each $H$. This implies immediately, that there are no quadratic invaraints and the quartic invariant is unique and given by $J_4^{(10d)}\sim\Sigma^{a_1 b_1, c_1 d_1}\Sigma^{a_2 b_2, c_2 d_2}H_{a_1 a_2}H_{b_1 b_2}H_{c_1 c_2}H_{d_1 d_2}$, equivalent to \eqref{I410d} up to a factor.

The uniqueness of quartic interactions of chiral forms in 10d has been established in a different language in \cite{Buratti:2019guq}. There are two advantages of our present treatment. First, this statement is translated into a straightforward, purely algebraic fact that there is a unique functionally independent quartic scalar that can be constructed from a chiral form in 10d. Second,
the unique independent quartic invariant (\ref{I410d}) alone yields an infinite-parametric family of interacting chiral 5-form theories by using an arbitrary function of this invariant in place of $\mathcal{F}(H+\s H)$ in (\ref{lagnl}), rather than only controlling the form of quartic interactions. This is, evidently, still a tiny subclass of all theories described by (\ref{lagnl}) where any other invariants could be used as well.

\nopagebreak
Scale-invariant theories in 10d, analogous to (\ref{confchiral6d}), can also be constructed. Besides the evident interaction term $\sqrt{I_4^{\mathrm{(10d)}}}$, one can now choose an arbitrary $\cal F$ in (\ref{lagnl}) that is a homogeneous function of degree 2 in its arguments, yielding a huge class of scale-invariant theories.

\section{Covariant equations of motion}\label{eoms}

The past literature on the subject is often permeated with the perception that Abelian interactions of chiral form fields are strongly constrained, and the search for consistent interactions is seen as an important outstanding problem. The physical equations of motion (\ref{FFXX}-\ref{defXF}) of our general interacting chiral theory cast this matter in a rather different light, since they look like fairly generic Lorentz-covariant equations of motion, without any need to satisfy sophisticated restrictions on the form of the nonlinear terms.

If one seeks a nonlinear generalization of the free selfduality relation $H=\s H$, a natural starting point is
\beq
\s\!H=\mathcal{G}(H).
\eeq
This, however, is a set of algebraic equations with respect to the components of $H$, one equation per component, and thus one generally does not expect nontrivial dynamics for generic $\cG$ and the description of the set of admissible $\cG$'s is not straightforward. It is more illuminating to resolve these algebraic equations as an expression for $H-\s H$ in terms of $H+\s H$, that is,
\beq\label{HGH}
H- \s H=\mathfrak{G}(H+\s H).
\eeq
Generically, there are still as many equations here as there are form components. However, if it happens that $\mathfrak{G}$ satisfies $\mathfrak{G}=-\s\mathfrak{G}$ for any value of the arguments, the number of equations reduces to one half of the number of components, leaving precisely the amount of freedom one needs to specify a chiral form. Thus, the condition for $\mathfrak{G}$ to be admissible is merely that its values are anti-selfdual. The most general equations of such a form are
\beq
H- \s H=\mathfrak{g}(H+\s H)-\s\mathfrak{g}(H+\s H),
\label{mostgen}
\eeq
and these are essentially the most general Lorentz-covariant equations one may imagine to describe interacting extensions of free chiral theories, before discussing any further physical constraints.

The equations of motion (\ref{FFXX}-\ref{defXF}) produced by our Lagrangian theories are surprisingly close
to the most general conceivable equations (\ref{mostgen}), and furthermore these equations automatically come out in the form (\ref{mostgen}), rather than (\ref{HGH}), when varying the Lagrangian and gauging away the auxiliary fields. The only constraint one has to impose is that $\mathfrak{g}$ is obtained by differentiating a scalar made of $H+\s H$ with respect to the form components, as expressed by (\ref{defXF}). This is a surpisingly weak condition on the form of the interactions, which turns out sufficient to have a full-fledged Lagrangian form of the theory. (The derivative 
of a scalar ${\cal F}(H+\star H)$ with respect to the selfdual argument $H+\star H$ is always anti-selfdual in Minkowski spacetime in those dimensions where chiral forms exist.)

General results on the status of equations of motion of the form (\ref{mostgen}) which cannot be represented as (\ref{FFXX}-\ref{defXF}) will be reported elsewhere \cite{ZOK}. Although our formalism does not provide a Lagrangian description for such equations, we cannot make a general pronouncement on their consistency. The situation is simple, however, in the six-dimensional case of section~\ref{chi6d}, where we can show that any equation of the form (\ref{mostgen}) can be recast as (\ref{FFXX}-\ref{defXF}), and thus (\ref{chiral6d}) provides a Lagrangian description to the most general conceivable equations of motion (\ref{mostgen}). A proof of this statement is given in Appendix~\ref{tens6d}. The strategy is to show that very few functionally independent chiral forms can be constructed by contracting the tensor indices of chiral forms in 6d, and those functionally independent chiral forms can be expressed as derivatives of the scalar invariant (\ref{I4d6}). The technology that goes into the proof is a minor extension of the analogous proof of the uniqueness of the chiral form invariant in 6d, presented in Appendix~\ref{6d invariant}.

We conclude with an amusing observation in two dimensions. If we relax the assumption that the right hand side of \eqref{mostgen} should be polynomial with respect to its argument, we find (details will be provided in \cite{ZOK}) that there is an equation in this  case, describing half a scalar degree of freedom, similarly to \eqref{FFXX} or \eqref{mostgen}, but for which \eqref{defXF} does not hold (this equation should be understood as component-by-component relations for vectors):
\begin{align}
    \partial_\mu\varphi -\epsilon_{\mu\nu}\partial^{\nu}\varphi=\frac{1}{\partial^{\mu}\varphi+\epsilon^{\mu\rho}\partial_\rho \varphi}\,.\label{exceptionalEq}
\end{align}
In reality, this is not two equations, but one, since both sides of the equation are anti-selfdual.
It is clear that the right-hand side of \eqref{exceptionalEq} cannot arise from a derivative of a scalar, since there is no scalar that can be constructed from $\partial^+_\mu\varphi\equiv (\partial_\mu+\epsilon_{\mu\nu}\partial^\nu)\varphi$.
Equation \eqref{exceptionalEq} is equivalent to
\begin{align}
    \partial_{\mu}\varphi\,\partial^\mu\varphi=1\,,\label{dfdf1}
\end{align}
which can be rewritten as:
\begin{align}
    \partial_{+}\varphi\,\partial_{-}\varphi=-1\,,\label{d+d-}
\end{align}
where $x^\pm=x^0\pm x^1$ are the light-cone coordinates in two dimensions. Curiously, this equation is the same as the duality-symmetry condition for the non-linear electrodynamics Lagrangian in four dimensions \eqref{4dDSL}, despite the completely unrelated physical interpretation. The field theory described by \eqref{exceptionalEq} or \eqref{dfdf1} is strongly coupled and symmetric with respect to exchanging $F-\star F$ and $F+\star F$. It cannot be understood as a continuous interacting deformation of the selfdual free scalar theory, or equivalently, the anti-selfdual free scalar theory. In $d = 2$, \eqref{exceptionalEq} is the only example of an equation \eqref{mostgen} that cannot be derived from the general Lagrangian \eqref{lagnl}. More generally, \eqref{exceptionalEq} is the only consistent (interacting half-scalar) equation of the type \eqref{mostgen} in two dimensions. We will report in more detail on this case, as well as cases with $d>2$, in \cite{ZOK}.

\section{Conclusions}\label{concl}

Building on top of our recent work \cite{democ}, and the earlier developments for free fields in  \cite{Mkrtchyan:2019opf,polynom}, we have provided a local, Lagrangian, manifestly Lorentz-covariant democratic description (\ref{democX}) for general self-interactions of Abelian $p$-forms, explicitly featuring electric and magnetic potentials  on equal footing. Additionally, in those dimensions where selfdual forms exists, this approach immediately leads to local, Lagrangian, manifestly Lorentz-covariant theories (\ref{lagnl}) of self-interacting chiral forms. The class of theories constructed in this fashion is considerably broader than what has been accessible to any past approaches in the literature.

Besides being local, Lagrangian and manifestly Lorentz-covariant, our formulation has a few further  distinctive advantages:
\begin{enumerate}

\item The large set of gauge symmetries necessary for eliminating the auxiliary Lagrangian fields and leaving only the desired dynamical content on-shell, is realized in a universal manner.  The  {\it expressions for the gauge transformations do not depend on the form of the interactions} one chooses to include in the theory. 

\item The {\it interaction terms in the Lagrangian}, expressed through a specific combination of dynamical and auxiliary fields, {\it are only constrained by Lorentz invariance}. One does not need to satisfy any extra requirements (for example, in the form of PDEs) in relation to the dependence of the interaction terms on the field variables.
\end{enumerate}

We feel that our results invite considerable re-evaluation of a number of perceptions commonly seen in the literature on the subject over the past decades. We draw the reader's attention in particular to the following important points:
\begin{enumerate}

\item A recurring motif in the past literature is that one should give up manifest Lorentz covariance
to deal with interacting chiral forms, in particular in 6d. Thus, the influential paper \cite{PSch} remarks that
``not only is there no manifestly Lorentz invariant action, but even the field
equation lacks manifest Lorentz invariance.'' This attitude reverberates through later articles on this subject, for example \cite{non5brane}, where a similar approach based on a 5d description is pursued for non-Abelian form fields.\footnote{Concurrently with these attitudes, covariant Lagrangian description of chiral form interactions has been developed within the PST approach starting with \cite{S5M}, see \cite{Buratti:2019guq} for a more recent update. Up to six dimensions, the PST approach covers the same range of theories as our formalism, but our formalism becomes significantly more powerful in higher dimensions. We mention additionally the alternative approach of \cite{Sen1,Sen2} due to Ashoke Sen, which is generally suitable for discussing chiral form interactions, but where the auxiliary fields decouple from the dynamics, rather than being exactly gauged away.} Our work casts the matter in a different light, since not only extremely simple, manifestly Lorentz-covariant equations of motion (\ref{FFXX}-\ref{defXF}) are available for interacting chiral theories, involving an arbitrary scalar function $\mathcal{F}$ of a selfdual form variable, but these equations of motion can also be derived from the manifestly Lorentz-invariant Lagrangian (\ref{lagnl}). Furthermore, as has been shown in section~\ref{eoms} and Appendix~\ref{tens6d}, in six spacetime dimensions, these equations in fact cover the entire class of Lorentz-covariant nonlinear selfduality relations (\ref{mostgen}), which are essentially the most general equations of motion (without derivatives of the field strength) one may propose for interacting deformations of free chiral form equations.

\item There is a persistent attitude in the literature that interactions of chiral forms are somehow difficult to construct, and the interaction terms must satisfy stringent consistency conditions. Thus, we read in  \cite{KSV} that ``the functionals of gauge field
strengths which determine non-linear selfduality conditions are
constructed order-by-order as perturbative series expansions in powers
of the field strength and in general their explicit form is unknown
except for the Born-Infeld-type actions and few other examples.''
In the Lagrangian formulation (\ref{lagnl}), the interaction term $\cal F$ is only constrained by the elementary requirement of Lorentz invariance, and otherwise completely arbitrary. There are no further conditions that need to be imposed on $\cal F$ so as to make our formalism work.
\end{enumerate}

We thereby proclaim that the problem of constructing self-interactions of Abelian chiral forms has been solved, as well as the problem of democratic Lagrangian description of self-interacting Abelian forms, and that the solution has turned out embarrassingly simple.

\section*{Acknowledgments}

We are grateful to Amihay Hanany and Arkady Tseytlin for discussions, to John Schwarz for correspondence, and to Dmitri Sorokin for comments on the manuscript.
OE has been supported by the CUniverse research promotion project (CUAASC) at Chulalongkorn University. KM is supported by the European Union's Horizon 2020 research and innovation programme under the Marie Sk\lslash odowska-Curie grant number 844265.

\appendix

\section{Conversion between single-field and democratic formulations}\label{app:conv}

When converting between single-field and democratic representation of nonlinear electrodynamics in section~\ref{singlef}, one needs to solve the $2\times 2$ system of algebraic equations (\ref{galphabeta}-\ref{Lambdas}). Additionally, since $\alpha$ and $\beta$ are derivatives of a single function per (\ref{lagdev}), and $g_1$ and $g_2$ are derivatives of a single function per (\ref{g1g2def}), they must satisfy compatibility relations $\del\al/\del s+\del\be/\del p=0$ and $\del g_1/\del\lambda_2=\del g_2/\del\lambda_1$. From an immediate inspection, it is not obvious that such compatibility conditions will be automatically satisfied, and more broadly, that one can indeed integrate the solutions of (\ref{galphabeta}-\ref{Lambdas}) to obtain the functions $g$ and $\cal L$ appearing in the corresponding Lagrangians. The purpose of this appendix is to show that it is possible under mild assumptions.

We will find it convenient to work with complex combinations of the pairs of invariants $(s,p)$, $(-\lambda_2,\lambda_1)$ and, correspondingly, the analysis will be expressed in the complex variable notation.

\subsection{Preliminaries}

Let $D\subset\mathbb{C}$ be a connected open region and let $\mu\in C^1(D,\mathbb{C})$. Consider the equation
\begin{equation}
\frac{\partial f}{\partial z}=\mu.\label{DiffEq}
\end{equation}
The following statements are very easy to prove, but we will formulate them as lemmas for convenient further reference.

\begin{lemma} Equation (\ref{DiffEq}) has real-valued solutions on $D$ if and only if
\begin{equation}
\frac{\partial\mu}{\partial\bar z}=\frac{\partial\bar\mu}{\partial z},\quad\forall z\in D.\label{Holonom}
\end{equation}
In conjunction with equation (\ref{DiffEq}), this condition is equivalent to
\begin{equation}
\frac{\partial f}{\partial\bar z}=\bar\mu.\label{Holonom1}
\end{equation}
\end{lemma}
\begin{proof} It is clear that any real-valued (differentiable) solution of (\ref{DiffEq}) is automatically in $C^2(D,\mathbb{R})$, and the equality of mixed derivatives immediately yields (\ref{Holonom}). The equivalence of (\ref{Holonom}) and (\ref{Holonom1}) for a solution of (\ref{DiffEq}) is straightforward. The existence of a real-valued solution when (\ref{Holonom}) is satisfied will be established below in Lemma \ref{ExistLemma}.
\end{proof}
For simplicity, the following lemma is formulated under the assumption that $D$ is star-shaped (i.e., convex along every line through a fixed point), so that integration is taken along straight lines. But this can be replaced with any other choice of integration curves between points with some reasonable effort.

\begin{lemma}\label{ExistLemma} Assume that (\ref{Holonom}) is satisfied and that the region $D$ is star-shaped with respect to $z_0\in D$. Then for every $C\in\mathbb{R}$ equation (\ref{DiffEq}) has a unique real-valued solution $f\in C^2(D,\mathbb{R})$ satisfying $f(z_0)=C$ given by
\begin{equation}
f(z)=\int\limits^z_{z_0}\mu(\xi)d\xi+\int\limits^{\bar z}_{\bar z_0}\overline{\mu(\overline{\xi})}d\xi+C=\int\limits^z_{z_0}\mu(\xi)d\xi+\overline{\int\limits^z_{z_0}\mu(\xi)d\xi}+C\label{Sol},\quad\forall z\in D,
\end{equation}
where integration is performed along a straight line.
\end{lemma}
\begin{proof}We first convince ourselves that (\ref{Sol}) is indeed a real-valued $C^2$ solution with $f(z_0)=C$. The uniqueness follows from the fact that the general solution to the homogeneous equation $\frac{\partial f}{\partial z}=0$ are functions antiholomorphic in $D$, and an antiholomorphic function is real-valued only if it is a constant.
\end{proof}

In the sequel we will meet equations of the form
\begin{equation}
\frac{\partial F}{\partial\xi}\Big|_{\xi=\mu^2z}=\frac{k}{\mu},\quad k\in\mathbb{R},\label{ImplicEq}
\end{equation}
where $F(\xi)$ is given and we want to imply certain properties about the implicitly defined function $\mu(z)$, if it exists.

\begin{proposition}\label{SymProp} Let $D\subset\mathbb{C}$ be an open region and $\mu\in C^1(D,\mathbb{C}\setminus\{0\})$. Let further $\tilde D\subset\mathbb{C}$ be another open region such that $\mu(z)^2z\in\tilde D$ for all $z\in D$. Suppose that $F\in C^2(\tilde D,\mathbb{R})$ such that $\det\mathrm{H}F(\mu(z)^2z)\neq0$ for all $z\in D$. If $\mu$ satisfies equation (\ref{ImplicEq}) for some $k\in\mathbb{R}$ then it automatically satisfies (\ref{Holonom}).
\end{proposition}
\begin{proof} Differentiating both sides of equation (\ref{ImplicEq}) we find
$$
\frac{\partial}{\partial z}\frac{\partial F}{\partial\xi}(\mu^2z)=\frac{\partial^2F}{\partial\xi^2}\cdot\left(\mu^2+2\mu z\frac{\partial\mu}{\partial z}\right)+\frac{\partial^2F}{\partial\xi\partial\bar\xi}\cdot2\bar\mu\bar z\frac{\partial\bar\mu}{\partial z}=-\frac{k}{\mu^2}\frac{\partial\mu}{\partial z}.
$$
Repeating this for $\frac{\partial}{\partial\bar z}$, and taking also the complex conjugate equations, we can write the matrix equation
\begin{equation}
A\cdot\mathrm{D}\mu=C,\label{ADa=C}
\end{equation}
$$
A=\begin{pmatrix}
\frac{\partial^2 F}{\partial\xi^2}\cdot2\mu z+\frac{k}{\mu^2} && \frac{\partial^2 F}{\partial\xi\partial\bar\xi}\cdot2\bar\mu\bar z\\
\frac{\partial^2 F}{\partial\xi\partial\bar\xi}\cdot2\mu z && \frac{\partial^2 F}{\partial\bar\xi^2}\cdot2\bar\mu\bar z+\frac{k}{\bar\mu^2}
\end{pmatrix},\quad\mathrm{D}\mu=\begin{pmatrix}
\frac{\partial\mu}{\partial z} && \frac{\partial\mu}{\partial\bar z}\\
\frac{\partial\bar\mu}{\partial z} && \frac{\partial\bar\mu}{\partial\bar z}
\end{pmatrix},
$$
$$
C=-\begin{pmatrix}
\frac{\partial^2 F}{\partial\xi^2}\cdot\mu^2 && \frac{\partial^2 F}{\partial\xi\partial\bar\xi}\cdot\bar\mu^2\\
\frac{\partial^2 F}{\partial\xi\partial\bar\xi}\cdot\mu^2 && \frac{\partial^2 F}{\partial\bar\xi^2}\cdot\bar\mu^2
\end{pmatrix}.
$$
Observe that
$$
\det C(z)=|\mu(z)|^4\det\mathrm{H}F(\mu^2(z)z)\neq0,\quad\forall z\in D,
$$
whence $\det A(z)\neq0$ and thus $\mathrm{D}\mu(z)$ is uniquely determined from $A\cdot\mathrm{D}\mu=C$ for all $z\in D$. Moreover, a careful computation shows that $C(z)\cdot A(z)^\top=A(z)\cdot C(z)^\top$ for all $z\in D$, which implies that $\mathrm{D}\mu(z)=\mathrm{D}\mu(z)^\top$, and the latter is equivalent to (\ref{Holonom}).
\end{proof}

Aside from special situations, equation (\ref{ImplicEq}) will define a $C^1$ function $\mu(z)$ at least locally. Let us make this statement more precise.

\begin{proposition}\label{ExistProp} Let $\tilde D\subset\mathbb{C}$ be an open region and $F\in C^2(\tilde D,\mathbb{R})$ such that $\det\mathrm{H}F(w)\neq0$ for all $w\in\tilde D$. Let $D\subset\mathbb{C}$ and $D'\subset\mathbb{C}\setminus\{0\}$ be open regions such that
$$
(\forall (z,\mu)\in D\times D')\quad w=\mu^2z\in\tilde D.
$$
Then $\exists U\subset D$ open and $f\in C^1(U,D')$ such that $\mu=f(z)$ solves equation (\ref{ImplicEq}).
\end{proposition}
\begin{proof} Follows immediately from (\ref{ADa=C}) in the proof of Proposition \ref{SymProp} and the Implicit Function Theorem.
\end{proof}

\subsection{Existence results for the conversion procedure}

In formula (\ref{spdef}), we see that $s$ is a `parity-even' square and $p$ is a `parity-odd' square. This motivates the introduction of the complex variable
$$
z=s+\imath\, p.
$$
On the other hand, from formula (\ref{lamH}) we see that $\lambda_2$ is the negative of a `parity-even' square, while $\lambda_1$ is an `parity-odd' square. Thus, we introduce the complex variable
$$
\lambda=-\lambda_2+\imath\,\lambda_1.
$$
Henceforth we will consider $\mathcal{L}$ as a function of the complex variable $z$, and $g$ as a function of the complex variable $\lambda$. Denote
$$
v=1+\beta+\imath\,\alpha.
$$
In these new variables, equation (\ref{lagdev}) can be written as
\begin{equation}
\frac{\partial\mathcal{L}}{\partial z}=\frac{v-1}2,\quad\frac{\partial\mathcal{L}}{\partial\bar z}=\frac{\bar v-1}2,\label{LEq}
\end{equation}
where the second equation can be substituted by the requirement that $\mathcal{L}$ is real-valued. Similarly, equation (\ref{galphabeta}) can be written as
\begin{equation}
\frac{\partial g}{\partial\lambda}=\frac1v-\frac12,\quad\frac{\partial g}{\partial\bar\lambda}=\frac1{\bar v}-\frac12,\label{gEq}
\end{equation}
where the second equation can be substituted by the requirement that $g$ is real-valued. The connecting equation (\ref{Lambdas}) can be now written as
\begin{equation}
\lambda=v^2z.\label{Conn}
\end{equation}
Note that (\ref{walphabetasp}) follows immediately.

The question at hand is therefore: given a real-valued function $\mathcal{L}$, find a real-valued function $g$ such that equations (\ref{LEq}), (\ref{gEq}) and (\ref{Conn}) are satisfied, and vice versa. If $\mathcal{L}(z)$ is given, then by (\ref{LEq}) and (\ref{Conn}) we have
\begin{equation}
v=1+2\frac{\partial\mathcal{L}}{\partial z}\left(\frac{\lambda}{v^2}\right),\label{EqL}
\end{equation}
which we need to solve in order to find the function $v(\lambda)$. Conversely, if $g(\lambda)$ is given, then by (\ref{gEq}) and (\ref{Conn}) we have
\begin{equation}
\frac2v=1+2\frac{\partial g}{\partial\lambda}(v^2z),\label{Eqg}
\end{equation}
which we need to solve in order to find the function $v(z)$.

We are ready for the main result of this appendix.

\begin{theorem} Given a real-valued $C^2$ function $\mathcal{L}$, in a neighborhood of a point where $\det\mathrm{H}\mathcal{L}\neq0$ there exists a real-valued $C^2$ function $g$ such that (\ref{lagdev}), (\ref{galphabeta}) and (\ref{Lambdas}) are satisfied. Conversely, given a real-valued $C^2$ function $g$, in a neighborhood of a point where $\det\mathrm{H}g\neq0$ there exists a real-valued $C^2$ function $\mathcal{L}$ such that (\ref{lagdev}), (\ref{galphabeta}) and (\ref{Lambdas}) are satisfied.
\end{theorem}
\begin{proof} Let $\tilde D\subset\mathbb{C}$ open and $\mathcal{L}\in C^2(\tilde D,\mathbb{R})$, and let $w_0\in\tilde D$ be such that $\det\mathrm{H}\mathcal{L}(w_0)\neq0$. Then by continuity $\exists U\subset\tilde D$ open such that $w_0\in U$ and $\det\mathrm{H}\mathcal{L}(w)\neq0$ for all $w\in U$. We will assume $U=\tilde D$ for convenience. Define $F\in C^2(\tilde D,\mathbb{R})$ by
$$
F(\xi)=\xi+\bar\xi+2\mathcal{L}(\xi),\quad\forall\xi\in\tilde D.
$$
One sees immediately that $\det\mathrm{H}F(w)=4\det\mathrm{H}\mathcal{L}(w)\neq0$ for all $w\in\tilde D$. With notations $\mu=\frac1v$, $z=\lambda$, $k=1$, equation (\ref{EqL}) is equivalent to (\ref{ImplicEq}), which by Proposition \ref{ExistProp} has a solution $\mu\in C^1(D,D')$ for appropriately chosen neighborhoods $D,D'\subset\mathbb{C}$. By Proposition \ref{SymProp} this solution satisfies $\frac{\partial\mu}{\partial\bar\lambda}=\frac{\partial\bar\mu}{\partial\lambda}$. It follows that
$$
\frac{\partial}{\partial\bar\lambda}\left(\frac1v-\frac12\right)=\frac{\partial\mu}{\partial\bar\lambda}=\frac{\partial\bar\mu}{\partial\lambda}=\frac{\partial}{\partial\lambda}\left(\frac1{\bar v}-\frac12\right),
$$
so that by (\ref{gEq}) and Lemma \ref{ExistLemma} there exists $g\in C^2(D,\mathbb{R})$ that satisfies (\ref{gEq}).

The reverse implication is proven completely analogously with notations
$$
F(\xi)=\xi+\bar\xi+2g(\xi),\quad\mu=v,\quad z=z,\quad k=2.
$$
This completes the proof. 
\end{proof}

\begin{remark} Note that $\det\mathrm{H}\mathcal{L}\neq0$ (respectively, $\det\mathrm{H}g\neq0$) is the least we can require in order that the equations of correspondence are guaranteed to be non-degenerate and locally solvable. If this condition is not satisfied, solutions may still exist but this becomes much harder to establish on general grounds and should be elaborated case-by-case. Moreover, if $\mathcal{L}$ or $g$ are to be interpreted as Lagrangians, non-degenerate Hessians are important for finding local minima. In particular, strictly convex functions have positive definite Hessians.
\end{remark}

\section{Uniqueness of chiral form invariants}

\subsection{Six spacetime dimensions}\label{6d invariant}

For a fully antisymmetric tensor, a general polynomial invariant is a contraction of an arbitrary number of copies of this tensor and Levi-Civita symbols $\eps_{j_1j_2j_3j_4j_5j_6}$. For a selfdual rank 3 fully antisymmetric tensor $Y$, all the Levi-Civita symbols can be eliminated by first taking one of the $Y$'s that the Levi-Civita symbol is contracted to and rewriting it as in (\ref{sddef}) and then eliminating a pair of Levi-Civita symbols using the standard identity
\beq
\eps^{i_1i_2i_3i_4i_5i_6}\eps_{j_1j_2j_3j_4j_5j_6}=-\,\mathrm{det}_{mn}[\de^{i_m}_{j_n}].
\label{epsid}
\eeq
The minus sign originates from the Minkowski signature of the metric. Once this process has been applied to all Levi-Civita symbols, an arbitrary invariant of $Y$ is expressed as a contraction of $Y$'s only. Evidently, this contraction must involve an even number of $Y$'s so that there is an even total number of indices available for the contraction process.

Our goal is to prove that, in six spacetime dimensions, any invariant of a selfdual fully antisymmetric rank 3 tensor $Y$ can be expressed through
\beq\label{defI4app}
I_4^{\mathrm{(6d)}}= Y_{ikl}Y^{jkl}Y^{imn}Y_{jmn}.
\eeq
This statement underlies the analysis of section~\ref{chi6d}.
The dramatic reduction in the number of independent invariants can be understood, in particular, as an 
effect of the large number of identities on contractions of $Y$ resulting from the selfduality property.
One such important identity results from starting with $Y_{ijk}Y^{ilm}$, re-expressing both $Y$'s 
using (\ref{sddef}), and then applying (\ref{epsid}) to the emerging pair of Levi-Civita symbols.\footnote{Such computations are rather burdensome for manual implementation, but they are handled very efficiently by the FORM computer algebra system \cite{form}, which is optimized for working with tensor contractions and Levi-Civita symbols.} The result of these manipulations is the following identity
\beq
Y_{ijk}Y^{ilm}=\frac14\left(M_{j}^{\,\,l}\de_k^{m} -M_{j}^{\,\,m}\de_k^{l} 
-M_{k}^{\,\,l}\de_j^{m} +M_{k}^{\,\,m}\de_j^{l}\right),
\label{FAid}
\eeq
where $M_i^{\,\,j}\equiv Y_{ikl}Y^{jkl}$, as in (\ref{Mdef}).
In other words, a single contraction of a given pair of $Y$'s can be traded for a double contraction, which produces an $M$, while the contractions of the remaining $Y$'s in the invariant under consideration get `re-wired' as a result.

There is a further identity we shall need. Consider
\beq
M_i^{\,\,j} M_{j}^{\,\, k}= Y_{ipq}Y^{jpq}Y_{jrs}Y^{krs},
\eeq
and apply (\ref{FAid}) to the middle pair of $Y$'s.
This yields
\beq
M_i^{\,\,j} M_{j}^{\,\, k}= \frac14\left(Y_{ipq}Y^{krq}M_{r}^{\,\,p}\ +Y_{iqp}Y^{krp}M_{r}^{\,\,q} 
+Y_{ipq}Y^{ksq}M_{s}^{\,\,p} +Y_{iqp}Y^{ksp}M_{s}^{\,\,q}\right)=Y_{ipq}Y^{krq}M_{r}^{\,\,p}.
\eeq
Now apply (\ref{FAid}) on the remaining pair of $Y$'s, which yields
\beq
M_i^{\,\,j} M_{j}^{\,\, k}= \frac14\left(M_{i}^{\,\,k}\de_p^{r} -M_{i}^{\,\,r}\de_p^{k} 
-M_{p}^{\,\,k}\de_i^{r} +M_{p}^{\,\,r}\de_i^{k}\right)M_{r}^{\,\,p}=-\frac12 M_i^{\,\,j} M_{j}^{\,\, k}+\frac{I_4^{\mathrm{(6d)}}}4\de_i^k,
\eeq
where we have used 
\beq
M_i^{\,\,i}=\Tr[M]=Y_{ijk}Y^{ijk}=0,
\label{trzero}
\eeq
which is simply $Y\w\s Y=Y\w Y=0$ in the differential form notation.
Hence,
\beq
M_i^{\,\,j} M_{j}^{\,\, k}=\frac{I_4^{\mathrm{(6d)}}}6\de_i^k,
\label{Mtr}
\eeq

We are now ready to present an algorithm to express any full contraction of an even number of $Y$'s (and hence any scalar made of $Y$) through powers of $I_4^{\mathrm{(6d)}}$ defined by (\ref{defI4}). Starting with an arbitrary such contraction, execute the following 3 steps repeatedly, until they cannot be applied any further:
\begin{enumerate}
\item Replace all pairs of $Y$ connected via double contractions by $M$ using (\ref{Mdef}).
\item Repeatedly apply (\ref{Mtr}) to any pairs of $M$ contracted to each other, which eliminates such pairs of $M$ and brings in explicit powers of $I_4^{\mathrm{(6d)}}$.
\item Choose any pair of $Y$'s connected via a single contraction and convert it into $M$ using (\ref{FAid}).
\end{enumerate}
When the above steps can no longer be executed, one is left with a scalar made of $Y$'s and $M$'s where no $Y$ is contracted to another $Y$ and no $M$ is contracted to another $M$, times a power of  $I_4^{\mathrm{(6d)}}$. If the way this happens is that there are no $Y$'s and no $M$'s left at all, we have expressed the original invariant as a power of $I_4^{\mathrm{(6d)}}$. If there is one $M$ left, it must be contracted with itself, which is zero by (\ref{trzero}). If there are any $Y$'s left, each of them must have all of its indices contracted to different $M$'s (since they cannot be either contracted to other $Y$'s, by the recursive application of steps 1-3 above, nor to each other, since $Y$ is antisymmeric, nor to the same $M$, since $M$ is symmetric and $Y$ is antisymmetric). Choose one of the remaining $Y$'s, which must appear in the combination
\beq
M_{i'}^{\,\,i} M_{j'}^{\,\,j} M_{k'}^{\,\,k}\, Y_{ijk}=M_{j'}^{\,\,j} M_{k'}^{\,\,k}\,Y_{i'lm}\,Y^{ilm} Y_{ijk}.
\eeq
Apply (\ref{FAid}) to the last pair of $Y$'s in this expression to obtain
\begin{align}
M_{i'}^{\,\,i} M_{j'}^{\,\,j} M_{k'}^{\,\,k}\, Y_{ijk}&=\frac14 M_{j'}^{\,\,j} M_{k'}^{\,\,k}\,Y_{i'lm}\left(M_{j}^{\,\,l}\de_k^{m} -M_{j}^{\,\,m}\de_k^{l}
-M_{k}^{\,\,l}\de_j^{m} +M_{k}^{\,\,m}\de_j^{l}\right)\nonumber\\ 
&=\frac{I_4^{\mathrm{(6d)}}}{12} \left(M_{k'}^{\,\,k}\,Y_{i'j'k} +M_{j'}^{\,\,j} \,Y_{i'jk'}\right),\label{YMMM}
\end{align}
where (\ref{Mtr}) has been used in the second line. Since index $i'$ originally belonged to an $M$, following the recursive implementation of steps 1-3 above, it had to be contracted to an $Y$ (and not to an $M$). After (\ref{YMMM}) has been enacted, this index is attached to an $Y$, so there are two $Y$'s contracted to each other, and we can restart the recursive application of steps 1-3 above. If there are any $Y$'s left after that, we shall apply (\ref{YMMM}) again, and restart 1-3, and so on. Since the number of $Y$'s constantly decreases in this process, repeated application of the above procedure must terminate with complete elimination of all $Y$'s. In that case, either there is one $M$ left, and it is contracted to itself and hence zero by (\ref{trzero}), or there are no tensors left at all, and the original scalar has been completely expressed as a power of $I_4^{\mathrm{(6d)}}$.

\subsection{Ten spacetime dimensions}\label{inv10d}

We turn to functionally independent scalars that can be made in 10 spacetime dimensions from a selfdual fully antisymmetric rank 5 tensor $Y$, and focus on quartic invariants. First of all, we introduce
\beq
M_i^{\,\,j}\equiv Y_{iklmn}Y^{jklmn},
\label{Mdef10}
\eeq
analogous to (\ref{Mdef}). Similarly to (\ref{FAid}), the following identity holds whenever $Y$ is selfdual:
\beq
Y^{ijpqr}Y_{mnpqr}=\frac18\left(M_m^{\,\,i}\de^j_n - M_n^{\,\,i}\de^j_m - M_m^{\,\,j}\de^i_n + M_n^{\,\,j}\de^i_m\right).
\label{FAid10}
\eeq
Then we can start constructing quartic invariants by distributing contractions between four copies of $Y$. Pick one of these four $Y$'s. The contractions of its 5 indices can be distributed among the three remaining copies of $Y$ as $4+1+0$ or $3+2+0$ or $3+1+1$ or $2+2+1$. If the pattern is $4+1+0$, there is only one way to complete the remaining contractions, and it results in the invariant
\beq
I_4^{\mathrm{(10d)}}= Y_{iklmn}Y^{jklmn}Y^{ipqrs}Y_{jpqrs},
\label{I410dapp}
\eeq 
analogous to (\ref{I4d6}). If the pattern is $3+2+0$, there is also only one way to complete the remaining contractions, giving $Y_{ijklm}Y^{pqklm}Y^{ijnrs}Y_{pqnrs}$. But then, applying (\ref{FAid10}) to the first pair of $Y$'s reduces this invariant to (\ref{I410d}). If the contraction pattern is $3+1+1$, the only way to complete the remaining contractions, giving $Y_{ijklm}Y^{pqklm}Y^{iq'nrs}Y_{pj'nrs}\,\eta_{qq'}\eta^{jj'}$. Again, applying (\ref{FAid10}) to the first pair of $Y$'s reduces this invariant to (\ref{I410d}). Finally, the only way to complete the remaining contractions in the $2+2+1$ pattern is
\beq
J_4^{\mathrm{(10d)}}= Y_{ijklm}Y^{ijnpq}Y^{rsn'lm}Y_{rsk'pq}\,\eta^{kk'}\eta_{jj'}.
\label{J410d}
\eeq 
We have verified numerically that the relation $J_4^{\mathrm{(10d)}}=I_4^{\mathrm{(10d)}}/6$ holds for any random initialization of the independent components of $Y$. The proof of this identity would use the analog of (\ref{FAid10}) where one starts with only one contraction of the two $Y$'s and expresses it through double and quadruple contractions. Then acting with this single-contraction identity on either of the two pairs of indices in (\ref{J410d}) contracted through the Minkowski metric $\eta$ will result in an expression for $J_4^{\mathrm{(10d)}}$ as a linear combination of itself and $I_4^{\mathrm{(10d)}}$, which gives the desired linear relation between the two quartic invariants. Implementing this process is rather cumbersome as the identities and index permutation counting involved become rather bulky, hence we see little merit in doing this here explicitly.

We conclude that all quartic invariants of a selfdual fully antisymmetric rank 5 tensor in 10d can be expressed through (\ref{I410dapp}), as claimed in section~\ref{chi10d}.

\section{Tensor functions of chiral forms in 6d}\label{tens6d}

We would like to analyze the general equations of motion (\ref{mostgen}) for a chiral form in six dimensions, and show that they can in fact be reduced to the form (\ref{FFXX}-\ref{defXF}), and hence derived from the Lagrangian (\ref{chiral6d}). To this end, we return to the notation of section~\ref{chi6d} and Appendix~\ref{6d invariant}, and introduce $Y\equiv H+\s H$ and $M$ defined by (\ref{Mdef}). In this language, $\mathfrak{g}(Y)$ of (\ref{mostgen}) is a fully antisymmetric rank 3 tensor made by contracting (a necessarily odd number of) $Y$'s. We would like to use the identities for chiral forms to reduce $\mathfrak{g}$ to $\del\mathcal{F}/\del Y$, where $\mathcal F$ is a polynomial function of the unique functionally independent invariant $I_4^{(6d)}$ defined by (\ref{defI4app}). 

The reduction algorithm described under (\ref{Mtr}) will still work, whereby any pair of contracted $Y$'s is expressed through $M$, and any pair of contracted $M$'s can be expressed through $I_4^{(6d)}$, and so on recursively, and if this procedure stalls, it can be restarted with (\ref{YMMM}). 
This procedure can be applied to any $Y$ that does not have open indices, so that any $Y$'s without open indices will be converted to powers of $I_4^{(6d)}$. One will then be left with $Y$'s that do have at least one uncontracted index, and furthermore cannot be contracted to each other directly, but only through $M$, since otherwise we would have been able to immediately apply (\ref{FAid}). Since there are only 3 open indices, at  most three $Y$'s can be left in this way. If these 3 indices are attached to three different $Y$'s, they must be contracted to form the combination
\beq
Y_{ilm}Y_{jnp}Y_{kqs}M^{ln}M^{pq}M^{ms},
\eeq
where the open indices $i$, $j$, $k$ may or may not be contracted to extra copies of $M$. Then, using the identity
\begin{align}\label{MMY}
M_i^{\,\,l}M_j^{\,\,m}Y_{lmk}&=M_i^{\,\,l}Y_{jpq}Y^{mpq}Y_{mkl}=
\frac14 M_i^{\,\,l}Y_{jpq}\left(M_{k}^{\,\,p}\de_{l}^{q}
 -M_{k}^{\,\,q}\de_{l}^{p} -M_{l}^{\,\,p}\de_k^{q} +M_{l}^{\,\,q}\de_k^{p}\right)\nonumber\\
&=\frac12 M_i^{\,\,q}M_{k}^{\,\,p}Y_{jpq}+\frac{I_4^{(6d)}}{12} Y_{ijk},
\end{align}
followed by (\ref{FAid}), two $Y$'s will be eliminated, leaving an object with one $Y$.

The situation where two of the open indices are attached to one $Y$ and the remaining one to the other $Y$ is impossible, since one would not be able to complete the contractions to form a rank 3 tensor. We are thus left with a situation where only one Y is left, and its open indices can be decorated with $M$'s in different ways, leaving four distinct cases (each of these tensor structures may be multiplied by an arbitrary function of the unique quartic scalar $I_4^{(6d)}$):
\begin{enumerate}
\item $Y_{ijk}$;
\item $M_i^{\,\,l}Y_{ljk}+M_j^{\,\,l}Y_{lki}+M_k^{\,\,l}Y_{lij}$;
\item $M_i^{\,\,l}M_j^{\,\,m}Y_{lmk}+M_j^{\,\,l}M_k^{\,\,m}Y_{lmi}+M_k^{\,\,l}M_i^{\,\,m}Y_{lmj}$;
\item $M_i^{\,\,l}M_j^{\,\,m}M_k^{\,\,n}Y_{lmn}$.
\end{enumerate}
Option 1 gives zero identically when substituted to the right-hand side of (\ref{mostgen}), and hence cannot contribute. Option 2 is precisely in the form of equations of motion of the chiral Lagrangian theory (\ref{chiral6d}). Option 4 is equivalent to option 2 by (\ref{YMMM}). Finally, for option 3, we can use
(\ref{MMY}) to express it through $Y_{ijk}$ times $I_4^{(6d)}$. It is thus effectively equivalent to option 1 and cannot contribute for the same reason. The bottom line is that any equations of motion of the form (\ref{mostgen}) in six dimensions can be re-written as equations of motion of the Lagrangian theory (\ref{chiral6d}). Thus, {\it the Lagrangian \eqref{chiral6d} covers all possible interacting deformations of a free chiral two-form theory in six dimensions}.

\end{document}